\documentclass[11pt]{article}%
\usepackage{amssymb}
\usepackage[singlespacing]{setspace}
\usepackage{amsmath}
\usepackage{amsfonts}
\usepackage{graphicx}
\usepackage{cite}%
\allowdisplaybreaks
\begin{document}

\author{\vspace{0.16in}Hartmut Wachter\thanks{e-mail:
Hartmut.Wachter@physik.uni-muenchen.de}\\Max-Planck-Institute\\for Mathematics in the Sciences\\Inselstr. 22, D-04103 Leipzig\\\hspace{0.4in}\\Arnold-Sommerfeld-Center\\Ludwig-Maximilians-Universit\"{a}t\\Theresienstr. 37, D-80333 M\"{u}nchen}
\title{Non-relativistic Schr\"{o}dinger theory on q-deformed quantum spaces I\\{\small Mathematical framework and equations of motion}}
\date{}
\maketitle

\begin{abstract}
\noindent The aim of these three papers (I, II, and III) is to develop a
q-deformed version of non-relativistic Schr\"{o}dinger theory. Paper I
introduces the fundamental mathematical and physical concepts. The braided
line and the three-dimensional q-deformed Euclidean space play the role of
position space. For both cases the algebraic framework is extended by a time
element. A short review of the elements of q-deformed analysis on the spaces
under consideration is given. The time evolution operator is introduced in a
consistent way and its basic properties are discussed. These reasonings are
continued by proposing q-deformed analogs of the Schr\"{o}dinger and the
Heisenberg picture.\newpage

\end{abstract}
\tableofcontents

\section{Introduction}

From the beginning of quantum field theory to the present day there is a great
hope that the occurrence of infinities in the formalism of quantum field
theory results from an incomplete description of space-time at very small
distances \cite{Schw}. Bohr and Heisenberg have been the first to suggest that
quantum field theories should be formulated on a space-time lattice, since it
would imply the existence of a smallest length \cite{Heis}. One of the
earliest and most serious attempts towards this goal was the concept of
'quantized space-time' due to Snyder \cite{Sny47, Yan47}. Of course there were
many other researchers who took up this idea over and over again (see for
example Refs. \cite{Fli48, Hill55, Das60, Gol63}) and each of the different
approaches has its own difficulties and advantages. However, with the arrival
of quantum groups and quantum spaces \cite{Ku83, Wor87, Dri85, Jim85, Drin86,
RFT90, Tak90, Man88, Maj95, ChDe96, KS97, Maj93-Int} in the eighties of the
last century a new, very promising method to discretize space and time seems
to be available \cite{FLW96, CW98}. The observation that it leads to very
realistic deformations of classical\ space-time symmetries nourishes the hope
for a new powerful\ regularization schema \cite{GKP96, MajReg, Oec99}.

In this paper attention is focused on q-deformed quantum spaces \cite{CSSW90,
PW90, SWZ91, Maj91, LWW97, Maj94-10}. (For other deformations of space-time
see Refs. \cite{Lu92, Cas93, Dob94, DFR95, ChDe95, ChKu04, Koch04}.)
Concretely, we deal with the so-called 'braided line' and the
three-dimensional q-deformed Euclidean space. The braided line can be viewed
as deformation of the set of real numbers, while the three-dimensional
q-deformed Euclidean space is nothing other than a non-commutative version of
the classical three dimensional Euclidean space. Essentially for us is the
fact that on both spaces differential calculi exist which are recognized as
q-analogs of classical translational symmetry \cite{WZ91, CSW91, Song92,
OSWZ92, Maj93-2}. The aim of our paper is to show that this algebraic
framework is suitable to formulate a q-deformed version of non-relativistic
Schr\"{o}dinger theory.

Towards this end we first extend the coordinate algebras of braided line and
three-dimensional q-deformed Euclidean space by a time element in such a way
that it perfectly fits into the existing algebraic structures. Section
\ref{AlgSet} is devoted to this task. Our considerations will show us that the
time element is central in the corresponding space-time algebras and is
decoupled from space coordinates. In Sec.\thinspace\ref{EleqAn}\ these results
are combined with those of our previous work in Refs. \cite{WW01, BW01, Wac02,
Wac03, Wac04, Wac05, qAn} to give a q-deformed version of analysis to the
extended quantum spaces of braided line and three-dimensional q-deformed
Euclidean space. In doing so, we present expressions for calculating products,
partial derivatives, integrals, exponentials, translations, and braided
products on the quantum spaces under consideration. We will see that due to
its special algebraic properties the time element behaves like a commutative
coordinate and the expressions for its derivatives, integrals, and so on are
given by the classical ones. For the space coordinates, however, the situation
is somewhat different, since their derivatives, integrals, and so on lead to
one- and three-dimensional versions of the well-known q-calculus \cite{Kac00}.
In this manner we obtain space-time structures in which space is discretized
but time is still continuous.

Our results so far will then be used to introduce time evolution operators in
a consistent way. Towards this end we start in Sec.\thinspace\ref{SecTimEvo}
from q-deformed Taylor rules for the quantum spaces under consideration.
Exploiting the algebraic properties of space and momentum variables we finally
regain the well-known expressions for the time evolution operator. In
Sec.\thinspace\ref{SHPic} this result will enable us to formulate q-deformed
versions of the Schr\"{o}dinger and the Heisenberg picture in considerable
analogy to the undeformed case. Finally, Sec.\thinspace\ref{SecCon} closes the
considerations so far by a conclusion before we will take them up in part II
of this paper.

\section{Algebraic set-up\label{AlgSet}}

This section is devoted to the algebras we are dealing throughout the paper,
i.e. the braided line and the q-deformed Euclidean space in three dimensions.
It is our aim to enlarge both algebras by a time element and to derive
commutation relations for the generators of the extended algebras.

\subsection{Extended braided line \label{ExtBraAlg}}

Let us recall that the algebraic properties of quantum groups and quantum
spaces are completely encoded in their R-matrices, for which we require to
satisfy the quantum Yang-Baxter equation \cite{Maj95, ChDe96, KS97}:%
\begin{equation}
\hat{R}_{12}\hat{R}_{23}\hat{R}_{12}=\hat{R}_{23}\hat{R}_{12}\hat{R}_{23}.
\end{equation}
If we want to extend the algebra of the braided line by a time element with
trivial braiding the R-matrix for this space should take the form%
\begin{equation}
\hat{R}_{kl}^{ij}=%
\begin{pmatrix}
1 & 0 & 0 & 0\\
0 & 0 & 1 & 0\\
0 & 1 & 0 & 0\\
0 & 0 & 0 & q
\end{pmatrix}
, \label{ExtRma}%
\end{equation}
where $q>1$ and $i,j,k,l\in\{0,1\}.$ Notice that rows and columns of the above
matrix are arranged in the order $00,$ $01,$ $10,$ and $11.$ We take the
convention that the indices $0$ and $1$ respectively correspond to a time and
a space coordinate. One can check that the above matrix indeed gives a
solution to the Yang-Baxter equation (I am very grateful to Alexander Schmidt
for doing this calculation with Mathematica.). Thus, we refer to it as the
R-matrix of the so-called \textit{extended braided line}.

Next, we would like to find the commutation relations for the generators of
the extended braided line. Towards this end we first determine the eigenvalues
of the new R-matrix in (\ref{ExtRma}). They take on the values $1,$ $-1,$ and
$q.$ The projectors onto the corresponding eigenspaces are then given by%
\begin{gather}
P_{+}=\frac{(\hat{R}-\text{Id})(\hat{R}-q\text{Id})}{2(1+q)}=%
\begin{pmatrix}
0 & 0 & 0 & 0\\
0 & 1/2 & -1/2 & 0\\
0 & -1/2 & 1/2 & 0\\
0 & 0 & 0 & 0
\end{pmatrix}
,\\
P_{-}=\frac{(\hat{R}+\text{Id})(\hat{R}-q\text{Id})}{2(1-q)}=%
\begin{pmatrix}
1 & 0 & 0 & 0\\
0 & 1/2 & 1/2 & 0\\
0 & 1/2 & 1/2 & 0\\
0 & 0 & 0 & 0
\end{pmatrix}
,\\
P_{0}=\frac{(\hat{R}+\text{Id})(\hat{R}-\text{Id})}{(q+1)(q-1)}=%
\begin{pmatrix}
0 & 0 & 0 & 0\\
0 & 0 & 0 & 0\\
0 & 0 & 0 & 0\\
0 & 0 & 0 & 1
\end{pmatrix}
.
\end{gather}

The projector $P_{-}$ can be recognized as q-analog of an antisymmetrizer. For
this reason, it determines the commutation relations for the extended braided
line, i.e.%
\begin{equation}
(P_{-})_{kl}^{ij}\,X^{k}X^{l}=0\quad\Rightarrow\quad X^{0}X^{1}=X^{1}X^{0},
\label{QuaSpaRel1dim}%
\end{equation}
where summation over repeated indices is understood. The other two projectors
lead us to the commutation relations of the exterior algebra of the braided
line, since the differentials have to fulfill%
\begin{equation}
(P_{+})_{kl}^{ij}\,dX^{k}dX^{l}=0,\quad(P_{0})_{kl}^{ij}\,dX^{k}dX^{l}=0,
\end{equation}
or, more concretely,%
\begin{equation}
dX^{0}dX^{0}=0,\quad dX^{1}dX^{1}=0,\quad dX^{0}dX^{1}=-dX^{1}dX^{0}.
\label{DiffBrai}%
\end{equation}
Let us note that the commutation relations of the differentials can
alternatively be written as%
\begin{equation}
dX^{i}dX^{j}=(P_{-})_{kl}^{ij}\,dX^{k}dX^{l}=-\hat{R}_{kl}^{ij}\,dX^{k}dX^{l},
\end{equation}
which implies%
\begin{equation}
X^{i}dX^{j}=\hat{R}_{kl}^{ij}\,dX^{k}X^{l}. \label{VerDiffKoor}%
\end{equation}

As next step we introduce partial derivatives by
\begin{equation}
d=dX^{i}\partial_{i}, \label{DefPartDer}%
\end{equation}
where the exterior derivative $d$ obeys the usual properties of nilpotency and
Leibniz rule, i.e.
\begin{align}
d^{2}  &  =0,\nonumber\\
d(fg)  &  =(df)g+f(dg). \label{ExtDerivN}%
\end{align}
From (\ref{ExtDerivN}) together with (\ref{VerDiffKoor}) it can be shown that
we have as Leibniz rules%
\begin{equation}
\partial_{i}X^{j}=\delta_{i}^{j}+\hat{R}_{il}^{jk}X^{l}\partial_{k},
\label{DifCalExt}%
\end{equation}
or, more explicitly,%
\begin{align}
\partial_{0}X^{0}  &  =1+X^{0}\partial_{0},\nonumber\\
\partial_{0}X^{1}  &  =X^{1}\partial_{0},\label{Diff1dimUn1}\\[0.1in]
\partial_{1}X^{0}  &  =X^{0}\partial_{1},\nonumber\\
\partial_{1}X^{1}  &  =1+qX^{1}\partial_{1}. \label{Diff1dimUn2}%
\end{align}
For the sake of completeness, it should be noted that partial derivatives
satisfy the same commutation relations as quantum space coordinates, i.e.%
\begin{equation}
\partial_{0}\partial_{1}=\partial_{1}\partial_{0}.
\end{equation}

Now, we would like to enrich the algebraic structure by adding a conjugation.
A consistent choice is given by%
\begin{equation}
\overline{X^{0}}=X^{0},\quad\overline{X^{1}}=X^{1},
\end{equation}
and%
\begin{equation}
\overline{\partial_{0}}=-\partial_{0},\quad\overline{\partial_{1}}%
=-\partial_{1}.
\end{equation}
Applying this conjugation to the relations in (\ref{Diff1dimUn1}) and
(\ref{Diff1dimUn2}) yields a second differential calculus. Its relations read%
\begin{align}
\hat{\partial}_{0}X^{0}  &  =1+X^{0}\hat{\partial}_{0},\nonumber\\
\hat{\partial}_{0}X^{1}  &  =X^{1}\hat{\partial}_{0}, \label{Diff1dim1}%
\\[0.1in]
\hat{\partial}_{1}X^{0}  &  =X^{0}\hat{\partial}_{1},\nonumber\\
\hat{\partial}_{1}X^{1}  &  =1+q^{-1}X^{1}\hat{\partial}_{1},
\label{Diff1dim2}%
\end{align}
where $\hat{\partial}_{0}\equiv\partial_{0}$ and $\hat{\partial}_{1}\equiv
q\partial_{1}.$ In analogy to (\ref{DifCalExt}) we have
\begin{equation}
\hat{\partial}_{i}X^{j}=\delta_{i}^{j}+(\hat{R}^{-1})_{il}^{jk}X^{l}%
\hat{\partial}_{k}.
\end{equation}

Last but not least, we would like to say a few words about the quantum group
coacting on the extended braided line. If we require for the coaction
\begin{equation}
\beta(X^{i})=T_{j}^{i}\otimes X^{j},
\end{equation}
to be compatible with the algebraic structure of the extended braided line the
quantum group generators have to be subject to the relations%
\begin{equation}
\hat{R}_{kl}^{ij}\,T_{m}^{k}T_{n}^{l}=T_{k}^{i}T_{l}^{j}\hat{R}_{mn}^{kl},
\end{equation}
from which it follows that%
\begin{equation}
T_{j}^{i}=%
\begin{pmatrix}
a & 0\\
0 & b
\end{pmatrix}
\qquad\text{with }ab=ba.
\end{equation}
If we have $\overline{a}=a$ and $\overline{b}=b$ the extended braided line
even becomes a comodule-$\ast$-algebra.

\subsection{Extended three-dimensional q-deformed Euclidean space}

As in the previous subsection we start our considerations from the R-matrix.
The R-matrix for the three-dimensional q-deformed Euclidean space extended by
a time element is of block-diagonal form. Its building-blocks read
\cite{LSW94}%
\begin{gather}%
\begin{tabular}
[c]{ccc}
& $++$ & $--$\\\cline{2-3}%
$++$ & \multicolumn{1}{|c}{$1$} & $0$\\
$--$ & \multicolumn{1}{|c}{$0$} & $1$%
\end{tabular}
\ ,\\[0.1in]%
\begin{tabular}
[c]{ccc}
& $+3$ & $3+$\\\cline{2-3}%
$+3$ & \multicolumn{1}{|c}{$0$} & $q^{-2}$\\
$3+$ & \multicolumn{1}{|c}{$q^{-2}$} & $q^{-2}\lambda\lambda_{+}$%
\end{tabular}
\ ,\\[0.1in]%
\begin{tabular}
[c]{ccc}
& $3-$ & $-3$\\\cline{2-3}%
$3-$ & \multicolumn{1}{|c}{$0$} & $q^{-2}$\\
$-3$ & \multicolumn{1}{|c}{$q^{-2}$} & $q^{-2}\lambda\lambda_{+}$%
\end{tabular}
\ ,\\[0.1in]%
\begin{tabular}
[c]{llll}
& $+-$ & $33$ & $-+$\\\cline{2-4}%
$+-$ & \multicolumn{1}{|l}{$0$} & $0$ & $q^{-4}$\\
$33$ & \multicolumn{1}{|l}{$0$} & $q^{-2}$ & $q^{-3}\lambda\lambda_{+}$\\
$-+$ & \multicolumn{1}{|l}{$q^{-4}$} & $q^{-3}\lambda\lambda_{+}$ &
$q^{-3}\lambda^{2}\lambda_{+}$%
\end{tabular}
\ ,\\[0.1in]%
\begin{tabular}
[c]{llllllll}
& $00$ & $0+$ & $03$ & $0-$ & $+0$ & $30$ & $-0$\\\cline{2-8}%
$00$ & \multicolumn{1}{|l}{$1$} & $0$ & $0$ & $0$ & $1$ & $0$ & $0$\\
$0+$ & \multicolumn{1}{|l}{$0$} & $0$ & $0$ & $0$ & $0$ & $1$ & $0$\\
$03$ & \multicolumn{1}{|l}{$0$} & $0$ & $0$ & $0$ & $0$ & $0$ & $1$\\
$0-$ & \multicolumn{1}{|l}{$0$} & $0$ & $0$ & $0$ & $0$ & $0$ & $0$\\
$+0$ & \multicolumn{1}{|l}{$0$} & $1$ & $0$ & $0$ & $0$ & $0$ & $0$\\
$30$ & \multicolumn{1}{|l}{$0$} & $0$ & $1$ & $0$ & $0$ & $0$ & $0$\\
$-0$ & \multicolumn{1}{|l}{$0$} & $0$ & $0$ & $1$ & $0$ & $0$ & $0$%
\end{tabular}
\ .
\end{gather}
Notice that space coordinates are labeled by $+,3,$ or $-$, while the index
$0$ refers to the time element.

To get commutation relations for the coordinates in space and time we need to
know the eigenvalues of the above R-matrix. They are given by $1,$ $-q^{-4},$
$q^{-6},$ and $-1.$ Thus, the projectors onto the corresponding irreducible
subspaces can be calculated from the identities%
\begin{align}
P_{+}  &  =\frac{(\hat{R}+q^{-4}\text{Id})(\hat{R}-q^{-6}\text{Id})(\hat
{R}+\text{Id})}{2(1+q^{-4})(1-q^{-6})},\\[0.1in]
P_{-}  &  =\frac{(\hat{R}-\text{Id})(\hat{R}-q^{-6}\text{Id})(\hat
{R}+\text{Id})}{(1+q^{-4})(q^{-4}+q^{-6})(1-q^{-4})},\\[0.1in]
P_{0}  &  =\frac{(\hat{R}-\text{Id})(\hat{R}+q^{-4}\text{Id})(\hat
{R}+\text{Id})}{(q^{-6}-1)(q^{-6}+q^{-4})(q^{-6}+1)},\\[0.1in]
P^{\prime}  &  =\frac{(\hat{R}-\text{Id})(\hat{R}+q^{-4}\text{Id})(\hat
{R}-q^{-6}\text{Id})}{2(q^{-4}-1)(1+q^{-6})}.
\end{align}

The projectors $P_{-}$ and $P^{\prime}$ lead us to the defining\ relations of
the extended three-dimensional q-deformed Euclidean space:%
\begin{equation}
(P_{-})_{kl}^{ij}\,X^{k}X^{l}=0,\qquad(P^{\prime})_{kl}^{ij}\,X^{k}X^{l}=0.
\end{equation}
Written out explicitly, these relations become
\begin{gather}
X^{0}X^{+}=X^{+}X^{0},\quad X^{0}X^{-}=X^{-}X^{0},\quad X^{0}X^{3}=X^{3}%
X^{0},\nonumber\\
X^{-}X^{3}=q^{2}X^{3}X^{-},\quad X^{3}X^{+}=q^{2}X^{+}X^{3},\nonumber\\
X^{-}X^{+}-X^{+}X^{-}=\lambda X^{3}X^{3},
\end{gather}
where $\lambda=q-q^{-1}.$ It should be mentioned that under transformations of
the quantum group $SU_{q}(2)$ the quantum space coordinates $X^{+},$ $X^{3},$
and $X^{-}$ behave like components of a three-vector, while the time
coordinate $X^{0}$ transforms like a scalar. This situation is in complete
analogy to the undeformed case.

The exterior algebra to the extended q-deformed Euclidean space in three
dimensions is defined by the relations%
\begin{equation}
(P_{+})_{kl}^{ij}\,dX^{k}dX^{l}=0,\qquad(P_{0})_{kl}^{ij}\,dX^{k}dX^{l}=0.
\label{ExtAlg3dim}%
\end{equation}
From these relations we obtain%
\begin{align}
dX^{A}dX^{B}  &  =-q^{4}(\hat{R}_{SO_{q}(3)})_{CD}^{AB}\,dX^{C}dX^{D}%
\nonumber\\
&  =-q^{-4}(\hat{R}_{SO_{q}(3)}^{-1})_{CD}^{AB}\,dX^{C}dX^{D},
\end{align}
and%
\begin{equation}
dX^{0}dX^{0}=0,\quad dX^{0}dX^{A}=-dX^{A}dX^{0},
\end{equation}
where $A,B\in\{+,3,-\}.\ $Notice that\ $\hat{R}_{SO_{q}(3)}$ stands for the
R-matrix of the three-dimensional q-deformed Euclidean space without a time
element. Furthermore, we took the convention that capital letters like $A,B,$
etc. run through $(+,3,-).$

The commutation relations between differentials of coordinates require that
the braiding between quantum space coordinates and their differentials takes
the form%
\begin{equation}
X^{A}dX^{B}=q^{4}(\hat{R}_{SO_{q}(3)})_{CD}^{AB}\,dX^{C}X^{D},
\label{BraidXDif1}%
\end{equation}
or, alternatively,%
\begin{equation}
X^{A}dX^{B}=q^{-4}(\hat{R}_{SO_{q}(3)}^{-1})_{CD}^{AB}\,dX^{C}X^{D}.
\label{BraidXDif2}%
\end{equation}
If the time element is involved we additionally have%
\begin{equation}
X^{0}dX^{A}=dX^{A}dX^{0},\quad X^{A}dX^{0}=dX^{0}X^{A},\quad X^{0}%
dX^{0}=dX^{0}X^{0}. \label{BraidX0}%
\end{equation}
We recommend Ref. \cite{MSW04} if the reader is unfamiliar with the reasonings
leading to the above relations.

Now, we have everything together to introduce partial derivatives. We start
from the Leibniz rules%
\begin{align}
dX^{A}.  &  =(dX^{A}).+X^{A}d.,\nonumber\\
dX^{0}.  &  =(dX^{0}).+X^{0}d., \label{LeibExtDer}%
\end{align}
where the point stands for an additional element. Then we substitute the
exterior derivative by $d=dX^{i}\partial_{i}$ and obtain%
\begin{align}
dX^{i}\partial_{i}X^{A}  &  =dX^{A}+X^{A}dX^{i}\partial_{i},\nonumber\\
dX^{i}\partial_{i}X^{0}  &  =dX^{0}+X^{0}dX^{i}\partial_{i}.
\end{align}
Using relations (\ref{BraidXDif1}) and (\ref{BraidX0}) we are able to switch
all differentials to the far left:%
\begin{align}
dX^{i}\partial_{i}X^{A}  &  =dX^{A}+dX^{0}X^{A}\partial_{0}+q^{4}(\hat
{R}_{SO_{q}(3)})_{CD}^{AB}\,dX^{C}X^{D}\partial_{B},\nonumber\\
dX^{i}\partial_{i}X^{0}  &  =dX^{0}+dX^{i}X^{0}\partial_{i}.
\end{align}
Since the differentials $dX^{i},$ $i\in\{+,3,0,-\},$ are linearly independent,
it follows from the above identities that%
\begin{align}
\partial_{B}X^{A}  &  =\delta_{B}^{A}+q^{4}(\hat{R}_{SO_{q}(3)})_{CD}%
^{AB}\,X^{C}\partial_{D},\nonumber\\
\partial_{0}X^{A}  &  =X^{A}\partial_{0},
\end{align}
and%
\begin{equation}
\partial_{0}X^{0}=1+X^{0}\partial_{0},\quad\partial_{A}X^{0}=X^{0}\partial
_{A}.
\end{equation}

If we apply (\ref{BraidXDif2}) instead of (\ref{BraidXDif1}) in the above
derivation, we arrive at a second differential calculus with relations%
\begin{align}
\hat{\partial}_{B}X^{A}  &  =\delta_{B}^{A}+q^{-4}(\hat{R}_{SO_{q}(3)}%
^{-1})_{CD}^{AB}\,X^{C}\hat{\partial}_{D},\nonumber\\
\hat{\partial}_{0}X^{A}  &  =X^{A}\hat{\partial}_{0},
\end{align}
and%
\begin{equation}
\hat{\partial}_{0}X^{0}=1+X^{0}\hat{\partial}_{0},\quad\hat{\partial}_{A}%
X^{0}=X^{0}\hat{\partial}_{A},
\end{equation}
where $\hat{\partial}_{A}\equiv q^{6}\partial_{A}$ for $A\in\{+,3,-\}$ and
$\hat{\partial}_{0}\equiv\partial_{0}.$ For the sake of completeness let us
note that partial derivatives now obey the same commutation relations\ as
covariant quantum space coordinates, i.e.%
\begin{gather}
\partial_{0}\partial_{+}=\partial_{+}\partial_{0},\quad\partial_{0}%
\partial_{-}=\partial_{-}\partial_{0},\quad\partial_{0}\partial_{3}%
=\partial_{3}\partial_{0},\nonumber\\
\partial_{+}\partial_{3}=q^{2}\partial_{3}\partial_{+},\quad\partial
_{3}\partial_{-}=q^{2}\partial_{-}\partial_{3},\nonumber\\
\partial_{+}\partial_{-}-\partial_{-}\partial_{+}=\lambda\partial_{3}%
\partial_{3}.
\end{gather}

It is rather instructive to write the Leibniz rules out. In doing so, we
obtain
\begin{align}
\partial_{+}X^{0}  &  =X^{0}\partial_{+},\nonumber\\
\partial_{+}X^{+}  &  =1+q^{4}X^{+}\partial_{+},\nonumber\\
\partial_{+}X^{3}  &  =q^{2}X^{3}\partial_{+},\nonumber\\
\partial_{+}X^{-}  &  =X^{-}\partial_{+},\label{Lei3dimExp1}\\[0.16in]
\partial_{3}X^{0}  &  =X^{0}\partial_{3},\nonumber\\
\partial_{3}X^{+}  &  =q^{2}X^{+}\partial_{3},\nonumber\\
\partial_{3}X^{3}  &  =1+q^{2}X^{3}\partial_{3}+q^{2}\lambda\lambda_{+}%
X^{+}\partial_{+},\nonumber\\
\partial_{3}X^{-}  &  =q^{2}X^{-}\partial_{3}+q\lambda\lambda_{+}X^{3}%
\partial_{+},\\[0.16in]
\partial_{-}X^{0}  &  =X^{0}\partial_{-},\nonumber\\
\partial_{-}X^{+}  &  =X^{+}\partial_{-},\nonumber\\
\partial_{-}X^{3}  &  =q^{2}X^{3}\partial_{-}+q\lambda\lambda_{+}X^{+}%
\partial_{3},\nonumber\\
\partial_{-}X^{-}  &  =1+q^{4}X^{-}\partial_{-}+q^{2}\lambda\lambda_{+}%
X^{3}\partial_{3}+q\lambda^{2}\lambda_{+}X^{+}\partial_{+},\\[0.16in]
\partial_{0}X^{0}  &  =1+X^{0}\partial_{0},\nonumber\\
\partial_{0}X^{+}  &  =X^{+}\partial_{0},\nonumber\\
\partial_{0}X^{3}  &  =X^{3}\partial_{0},\nonumber\\
\partial_{0}X^{-}  &  =X^{-}\partial_{0}.
\end{align}
and%
\begin{align}
\hat{\partial}_{+}X^{0}  &  =X^{0}\hat{\partial}_{+},\nonumber\\
\hat{\partial}_{+}X^{-}  &  =X^{-}\hat{\partial}_{+},\nonumber\\
\hat{\partial}_{+}X^{3}  &  =q^{-2}X^{3}\hat{\partial}_{+}-q\lambda\lambda
_{+}X^{-}\hat{\partial}_{3},\nonumber\\
\hat{\partial}_{+}X^{+}  &  =1+q^{-4}X^{+}\hat{\partial}_{+}-q^{-2}%
\lambda\lambda_{+}X^{3}\hat{\partial}_{3}+q^{-1}\lambda^{2}\lambda_{+}%
X^{-}\hat{\partial}_{-},\\[0.16in]
\hat{\partial}_{3}X^{0}  &  =X^{0}\hat{\partial}_{3},\nonumber\\
\hat{\partial}_{3}X^{-}  &  =q^{-2}X^{-}\hat{\partial}_{3},\nonumber\\
\hat{\partial}_{3}X^{3}  &  =1+q^{-2}X^{3}\hat{\partial}_{3}-q^{-2}%
\lambda\lambda_{+}X^{-}\hat{\partial}_{-},\nonumber\\
\hat{\partial}_{3}X^{+}  &  =q^{-2}X^{+}\hat{\partial}_{3}-q^{-1}%
\lambda\lambda_{+}X^{3}\hat{\partial}_{-},\\[0.16in]
\hat{\partial}_{-}X^{0}  &  =X^{0}\hat{\partial}_{-}\nonumber\\
\hat{\partial}_{-}X^{+}  &  =X^{+}\hat{\partial}_{-},\nonumber\\
\hat{\partial}_{-}X^{3}  &  =q^{-2}X^{3}\hat{\partial}_{-},\nonumber\\
\hat{\partial}_{-}X^{-}  &  =1+q^{-4}X^{-}\hat{\partial}_{-},\\[0.16in]
\hat{\partial}_{0}X^{0}  &  =1+X^{0}\hat{\partial}_{0},\nonumber\\
\hat{\partial}_{0}X^{+}  &  =X^{+}\hat{\partial}_{3},\nonumber\\
\hat{\partial}_{0}X^{3}  &  =X^{3}\hat{\partial}_{0},\nonumber\\
\hat{\partial}_{0}X^{-}  &  =X^{-}\hat{\partial}, \label{Lei3dimExp2}%
\end{align}
where we set, for brevity, $\lambda_{+}\equiv q+q^{-1}.$

It remains to introduce a conjugation being compatible with the algebraic
structure presented so far. To this end, we need the quantum metric of the
three-dimensional q-deformed Euclidean space. Its explicit form can be read
off from the projector $P_{0},$ as it holds \cite{LWW97}%
\begin{equation}
(P_{0})_{CD}^{AB}=\frac{1}{g^{EF}g_{EF}}g^{AB}g_{CD}. \label{MetBes}%
\end{equation}
In this manner, one can verify that the non-vanishing entries of $g_{AB}$ and
$g^{AB}$ are given by
\begin{equation}
g^{+-}=g_{+-}=-q,\quad g^{-+}=g_{-+}=-q^{-1},\quad g^{33}=g_{33}=1.
\end{equation}
With the three-dimensional quantum metric at hand we are able to write down
the conjugation properties of coordinates and partial derivatives in a rather
compact form:%
\begin{equation}
\overline{X^{A}}=X_{A}=g_{AB}X^{B},\quad\overline{\partial_{A}}=-\partial
^{A}=-g^{AB}\partial_{B},
\end{equation}
and%
\begin{equation}
\overline{X^{0}}=X^{0},\quad\overline{\partial_{0}}=-\partial_{0}.
\end{equation}

\section{Elements of q-analysis\label{EleqAn}}

In this section a q-deformed version of analysis is given to the extended
braided line and the extended q-deformed Euclidean space in three dimensions.
Especially, we present expressions for computing star products, braided
products, q-translations, operator representations of partial derivatives,
q-integrals, and q-exponentials. With this toolbox of essential elements of
q-analysis we are in a position to formulate quantum mechanics on the quantum
spaces under consideration. Finally, it should be noted that the reasonings in
this section are mainly based on the ideas developed in Refs. \cite{WW01,
BW01, Wac02, Wac03, Wac04, Wac05, qAn, Maj93-Int, Maj93-5, CSW93, Schir94,
CHMW99, Maj94-10, Maj95, Maj95star}.

\subsection{Extended braided line\label{QAnBraid}}

First of all, let us mention that the product on the extended braided line is
the commutative one. This observation follows from a short look at the
commutation relations in Eq. (\ref{QuaSpaRel1dim}). However, if we want to
commute functions living in distinct quantum spaces things become slightly
different. In this respect, let us recall that the commutation relations
between generators of different quantum spaces are determined by the R-matrix
or its inverse, i.e.%
\begin{equation}
X^{i}\odot_{\bar{L}}Y^{j}\equiv(1\otimes X^{i})(Y^{j}\otimes1)=\hat{R}%
_{kl}^{ij}Y^{k}\otimes X^{l},
\end{equation}
and alternatively%
\begin{equation}
X^{i}\odot_{L}Y^{j}\equiv(1\otimes X^{i})(Y^{j}\otimes1)=(\hat{R}^{-1}%
)_{kl}^{ij}Y^{k}\otimes X^{l}.
\end{equation}
These relations lead us to braided products for commutative functions,%
\begin{align}
f(x^{i})\odot_{\bar{L}}g(y^{j})  &  =q^{\hat{n}_{y^{1}}\otimes\,\hat{n}%
_{x^{1}}}g(y^{j})\otimes f(x^{i}),\nonumber\\
f(x^{i})\odot_{L}g(y^{j})  &  =q^{-\hat{n}_{y^{1}}\otimes\,\hat{n}_{x^{1}}%
}g(y^{j})\otimes f(x^{i}),
\end{align}
where we introduced the operator%
\begin{equation}
\hat{n}_{x^{1}}\equiv x^{1}\frac{\partial}{\partial x^{1}}. \label{N-Op}%
\end{equation}
Notice that the partial derivative in Eq. (\ref{N-Op}) is a commutative one.
At this place it should also be mentioned that throughout this paper we use
the convention to write generators of quantum spaces in capitals, while
commutative coordinates are written in small letters. (In the case of the
braided line the identification with a commutative algebra is rather trivial.
For q-deformed Euclidean space in three dimensions, however, such an
identification needs some more thoughts.)

Next, we come to translations on the extended braided line. Translations on
quantum spaces are described by their Hopf structures. On quantum space
generators these Hopf structures become%
\begin{align}
\Delta_{L}(X^{0})  &  =X^{0}\otimes1+1\otimes X^{0},\nonumber\\
\Delta_{L}(X^{1})  &  =X^{1}\otimes1+\Lambda^{-1}\otimes X^{1}%
,\nonumber\\[0.1in]
S_{L}(X^{0})  &  =-X^{0},\nonumber\\
S_{L}(X^{1})  &  =-\Lambda X^{1},\nonumber\\[0.1in]
\epsilon_{L}(X^{i})  &  =0, \label{HopStrBrai1}%
\end{align}
and
\begin{align}
\Delta_{\bar{L}}(X^{0})  &  =X^{0}\otimes1+1\otimes X^{0},\nonumber\\
\Delta_{\bar{L}}(X^{1})  &  =X^{1}\otimes1+\Lambda\otimes X^{1}%
,\nonumber\\[0.1in]
S_{\bar{L}}(X^{0})  &  =-X^{0},\nonumber\\
S_{\bar{L}}(X^{1})  &  =-\Lambda^{-1}X^{1},\nonumber\\[0.1in]
\epsilon_{\bar{L}}(X^{i})  &  =0, \label{HopStrBrai2}%
\end{align}
where $\Lambda$ stands for a unitary scaling operator subject to%
\begin{gather}
\Lambda X^{0}=X^{0}\Lambda,\quad\Lambda X^{1}=qX^{1}\Lambda,\nonumber\\
\Lambda\partial_{0}=\Lambda\partial_{0},\quad\Lambda\partial_{1}=q^{-1}%
\Lambda\partial_{1}. \label{ActLam}%
\end{gather}
This scaling operator and its inverse can be viewed as generators of a Hopf
algebra denoted by $\mathcal{H}$. The corresponding Hopf structure reads%
\begin{equation}
\Delta(\Lambda)=\Lambda\otimes\Lambda,\quad S(\Lambda)=\Lambda^{-1}%
,\quad\epsilon(\Lambda)=1.
\end{equation}

To proceed any further we need the algebra morphisms $\mathcal{W}_{L}^{-1}$
and $\mathcal{W}_{R}^{-1}$ defined by
\begin{align}
\mathcal{W}_{L}^{-1}  &  :\mathcal{A}_{q}\rtimes\mathcal{H}\longrightarrow
\mathcal{A}_{q},\nonumber\\
\mathcal{W}_{L}^{-1}((X^{0})^{n_{0}}(X^{1})^{n_{1}}\otimes h)  &  \equiv
(x^{0})^{n_{0}}(x^{1})^{n_{1}}\,\varepsilon(h),\\[0.1in]
\mathcal{W}_{R}^{-1}  &  :\mathcal{H}\ltimes\mathcal{A}_{q}\longrightarrow
\mathcal{A}_{q},\nonumber\\
\mathcal{W}_{R}^{-1}(h\otimes(X^{0})^{n_{0}}(X^{1})^{n_{1}})  &
\equiv\varepsilon(h)\,(x^{0})^{n_{0}}(x^{1})^{n_{1}}.
\end{align}
With these mappings at hand we are able to introduce the operations%
\begin{align}
f(x^{i}\oplus_{L/\bar{L}}y^{j})  &  \equiv((\mathcal{W}_{L}^{-1}%
\otimes\mathcal{W}_{L}^{-1})\circ\Delta_{L/\bar{L}})(f),\\[0.16in]
f(\ominus_{L/\bar{L}}\,x^{i})  &  \equiv(\mathcal{W}_{R}^{-1}\circ
S_{L/\bar{L}})(f).
\end{align}
Repeating the same steps already applied in Ref. \cite{Wac04} one can show
that%
\begin{align}
f(x^{i}\oplus_{L}y^{j})  &  =\sum_{k,l=0}^{\infty}\frac{(x^{0})^{k}(x^{1}%
)^{l}}{k!\,[[l]]_{q^{-1}}!}\Big (\frac{\partial}{\partial y^{0}}%
\Big )^{k}(D_{q^{-1}}^{1})^{l}f(y^{j}),\label{TransL}\\[0.16in]
f(x^{i}\oplus_{\bar{L}}y^{j})  &  =\sum_{k,l=0}^{\infty}\frac{(x^{0}%
)^{k}(x^{1})^{l}}{k!\,[[l]]_{q}!}\Big (\frac{\partial}{\partial y^{0}%
}\Big )^{k}(D_{q}^{1})^{l}f(y^{j}), \label{TransLq}%
\end{align}
and%
\begin{align}
f(\ominus_{L}\,x^{i})  &  =q^{-\frac{1}{2}\hat{n}_{x^{1}}(\hat{n}_{x^{1}}%
-1)}f(-x^{i}),\\[0.16in]
f(\ominus_{\bar{L}}\,x^{i})  &  =q^{\frac{1}{2}\hat{n}_{x^{1}}(\hat{n}_{x^{1}%
}-1)}f(-x^{i}).
\end{align}
Notice that the expressions in (\ref{TransL}) and (\ref{TransLq}) use the
so-called \textit{Jackson derivatives} \cite{Jack08}%
\begin{equation}
D_{q^{a}}^{i}f\equiv\frac{f(q^{a}x^{i})-f\left(  x^{i}\right)  }%
{(q^{a}-1)x^{i}},\qquad a\in\mathbb{C}.
\end{equation}
Furthermore, the so-called antisymmetric q-numbers are given by%
\begin{equation}
\left[  \left[  n\right]  \right]  _{q^{a}}\equiv\sum_{k=0}^{n-1}q^{ak}%
=\frac{1-q^{an}}{1-q^{a}},
\end{equation}
and their factorials are defined by%
\begin{equation}
\left[  \left[  n\right]  \right]  _{q^{a}}!\equiv\left[  \left[  1\right]
\right]  _{q^{a}}\left[  \left[  2\right]  \right]  _{q^{a}}\ldots\left[
\left[  n\right]  \right]  _{q^{a}},\qquad\left[  \left[  0\right]  \right]
_{q^{a}}!\equiv1.
\end{equation}

Now, we would like to turn our attention to operator representations of
partial derivatives. From the q-deformed Leibniz rules in (\ref{Diff1dimUn1})
and (\ref{Diff1dimUn2}) as well as those in (\ref{Diff1dim1}) and
(\ref{Diff1dim2}) we can derive right and left actions of partial derivatives
on the algebra of quantum space coordinates. To this end, we repeatedly apply
the Leibniz rules to the product of a partial derivative with a normally
ordered monomial of coordinates, until we obtain an expression with all
partial derivatives standing to the right of all quantum space coordinates,
i.e.%
\begin{equation}
\partial^{i}(X^{0})^{k_{0}}(X^{1})^{k_{1}}=\big (\partial_{(1)}^{i}%
\triangleright(X^{0})^{k_{0}}(X^{1})^{k_{1}}\big )\partial_{(2)}^{i}.
\label{VerParX}%
\end{equation}
Taking the counit of all partial derivatives appearing on the right-hand side
finally yields the left action of $\partial^{i}$, since we have%
\begin{equation}
\big (\partial_{(1)}^{i}\triangleright(X^{0})^{k_{0}}(X^{1})^{k_{1}%
}\big )\varepsilon(\partial_{(2)}^{i})=\partial^{i}\triangleright
(X^{0})^{k_{0}}(X^{1})^{k_{1}}. \label{BerWirkPar}%
\end{equation}
Right actions of partial derivatives can be calculated in a similar way if we
start from a partial derivative standing to the right of a normally ordered
monomial and commute it to the left of all quantum space coordinates. Instead
of (\ref{VerParX}) and (\ref{BerWirkPar}) we have
\begin{equation}
(X^{0})^{k_{0}}(X^{1})^{k_{1}}\partial^{i}=\partial_{(2)}^{i}\big ((X^{0}%
)^{k_{0}}(X^{1})^{k_{1}}\triangleleft\partial_{(1)}^{i}\big ),
\end{equation}
and%
\begin{equation}
\varepsilon(\partial_{(2)}^{i})\big ((X^{0})^{k_{0}}(X^{1})^{k_{1}%
}\triangleleft\partial_{(1)}^{i}\big )=(X^{0})^{k_{0}}(X^{1})^{k_{1}%
}\triangleleft\partial^{i}.
\end{equation}

These reasonings show us a method to calculate explicit formulae for the
action of partial derivatives on normally ordered monomials. From these
results we can finally read off the operator representations
\begin{align}
\partial_{0}\triangleright f(x^{i})  &  =\frac{\partial}{\partial x^{0}%
}f(x^{i}),\nonumber\\
\partial_{1}\triangleright f(x^{i})  &  =D_{q}^{1}\,f(x^{i}%
),\label{ActPartBrai1}\\[0.16in]
\hat{\partial}_{0}\,\bar{\triangleright}\,f(x^{i})  &  =\frac{\partial
}{\partial x^{0}}f(x^{i}),\nonumber\\
\hat{\partial}_{1}\,\bar{\triangleright}\,f(x^{i})  &  =D_{q^{-1}}%
^{1}\,f(x^{i}),
\end{align}
and%
\begin{align}
f(x^{i})\triangleleft\hat{\partial}_{0}  &  =-\frac{\partial}{\partial x^{0}%
}f(x^{i}),\nonumber\\
f(x^{i})\triangleleft\hat{\partial}_{x}  &  =-D_{q^{-1}}^{1}\,f(x^{i}%
),\\[0.16in]
f(x^{i})\,\bar{\triangleleft}\,\partial_{0}  &  =-\frac{\partial}{\partial
x^{0}}f(x^{i}),\nonumber\\
f(x^{i})\,\bar{\triangleleft}\,\partial_{1}  &  =-D_{q}^{1}\,f(x^{i}).
\label{ActPartBraid2}%
\end{align}
With these formulae at hand it follows at once that
\begin{gather}
df(x^{i})=dx^{j}\partial_{j}\triangleright f(x^{i})=0\nonumber\\
\Leftrightarrow f(x^{i})\big |_{x^{0}=\,a}=f(x^{i})\big |_{x^{0}=\,b},\quad
f(x^{i})\big |_{x^{1}=\,a}=f(x^{i})\big |_{x^{1}=\,qa},
\end{gather}
for all $a,b\in\mathbb{C}$. Notice that the above condition characterizes
functions being constant from the point of view of q-deformation.

Next, we come to integrals on the extended braided line. (For the different
approaches to introduce integrals on q-deformed spaces see also Refs.
\cite{Wac02, qAn, Fio93, Sta96, Chry96, KM94, CSW93, WZ91, CHMW99}.) Integrals
can be recognized as operations being inverse to partial derivatives. Thus, we
first try to extend the algebra of partial derivatives by introducing inverse
elements. In doing so, we get as additional relations%
\begin{align}
(\partial_{0})^{-1}\partial_{0}  &  =\partial_{0}(\partial_{0})^{-1}%
=1,\nonumber\\
(\partial_{1})^{-1}\partial_{1}  &  =\partial_{1}(\partial_{1})^{-1}%
=1,\nonumber\\
(\partial_{0})^{-1}\partial_{1}  &  =\partial_{1}(\partial_{0})^{-1}%
,\nonumber\\
(\partial_{1})^{-1}\partial_{0}  &  =\partial_{0}(\partial_{1})^{-1}%
,\nonumber\\
(\partial_{0})^{-1}(\partial_{1})^{-1}  &  =(\partial_{1})^{-1}(\partial
_{0})^{-1}.
\end{align}

As next step we search for representations of the inverse partial derivatives
that fulfill the above relations. It should be obvious that they are given by
\begin{align}
(\partial_{0})^{-1}\big |_{x^{0}=a}^{b}\triangleright f(x^{i})  &  =\int
_{a}^{b}dx^{0}\,f(x^{i}),\nonumber\\
(\partial_{1})^{-1}\big |_{x^{1}=\,a}^{b}\triangleright f(x^{i})  &
=(D_{q}^{1})^{-1}\big |_{x^{1}=\,a}^{b}f(x^{i}),\\[0.16in]
(\hat{\partial}_{0})^{-1}\big |_{x^{0}=a}^{b}\,\bar{\triangleright}\,f(x^{i})
&  =\int_{a}^{b}dx^{0}\,f(x^{i}),\nonumber\\
(\hat{\partial}_{1})^{-1}\big |_{x^{1}=\,a}^{b}\,\bar{\triangleright}%
\,f(x^{i})  &  =(D_{q^{-1}}^{1})^{-1}\big |_{x^{1}=\,a}^{b}f(x^{i}),
\end{align}
and%
\begin{align}
f(x^{i})\triangleleft(\hat{\partial}_{0})^{-1}\big |_{x^{0}=\,a}^{b}  &
=-\int_{a}^{b}dx^{0}\,f(x^{i}),\nonumber\\
f(x^{i})\triangleleft(\hat{\partial}_{1})^{-1}\big |_{x^{1}=\,a}^{b}  &
=-(D_{q^{-1}}^{1})^{-1}\big |_{x^{1}=\,a}^{b}f(x^{i}),\\[0.16in]
f(x^{i})\,\bar{\triangleleft}\,(\partial_{0})^{-1}\big |_{x^{0}=\,a}^{b}  &
=-\int_{a}^{b}dx^{0}\,f(x^{i}),\nonumber\\
f(x^{i})\,\bar{\triangleleft}\,(\partial_{1})^{-1}\big |_{x^{1}=\,a}^{b}  &
=-(D_{q}^{1})^{-1}\big |_{x^{1}=\,a}^{b}f(x^{i}),
\end{align}
where $(D_{q}^{i})^{-1}$ denotes the \textit{Jackson integral} operator
\cite{Jack27}. For the sake of completeness we would like to give the
definition of the Jackson integral. For $a>0,$ $q>1,$ and $x^{i}>0,$ it
becomes
\begin{align}
(D_{q^{a}}^{i})^{-1}\big |_{0}^{x^{i}}f  &  =-(1-q^{a})\sum_{k=1}^{\infty
}(q^{-ak}x^{i})f(q^{-ak}x^{i}),\nonumber\\
(D_{q^{a}}^{i})^{-1}\big |_{x^{i}}^{\infty}f  &  =-(1-q^{a})\sum_{k=0}%
^{\infty}(q^{ak}x^{i})f(q^{ak}x^{i}),\nonumber\\
(D_{q^{-a}}^{i})^{-1}\big |_{0}^{x^{i}}f  &  =(1-q^{-a})\sum_{k=0}^{\infty
}(q^{-ak}x^{i})f(q^{-ak}x^{i}),\nonumber\\
(D_{q^{-a}}^{i})^{-1}\big |_{x^{i}}^{\infty}f  &  =(1-q^{-a})\sum
_{k=1}^{\infty}(q^{ak}x^{i})f(q^{ak}x^{i}), \label{Jackson1N}%
\end{align}
and, likewise for $a>0,$ $q>1,$ and $x^{i}<0,$
\begin{align}
(D_{q^{a}}^{i})^{-1}\big |_{x^{i}}^{0}f  &  =(1-q^{a})\sum_{k=1}^{\infty
}(q^{-ak}x^{i})f(q^{-ak}x^{i}),\nonumber\\
(D_{q^{a}}^{i})^{-1}\big |_{-\infty}^{x^{i}}f  &  =(1-q^{a})\sum_{k=0}%
^{\infty}(q^{ak}x^{i})f(q^{ak}x^{i}),\nonumber\\
(D_{q^{-a}}^{i})^{-1}\big |_{x^{i}}^{0}f  &  =-(1-q^{-a})\sum_{k=0}^{\infty
}(q^{-ak}x^{i})f(q^{-ak}x^{i}),\nonumber\\
(D_{q^{-a}}^{i})^{-1}\big |_{-\infty}^{x^{i}}f  &  =-(1-q^{-a})\sum
_{k=1}^{\infty}(q^{ak}x^{i})f(q^{ak}x^{i}). \label{Jackson2N}%
\end{align}

In analogy to the undeformed case we have rules for integration by parts. To
derive them we start from the Leibniz rules for partial derivatives. These
Leibniz rules can be read off from the coproducts in (\ref{HopStrBrai1}) and
(\ref{HopStrBrai2}). In this manner, we find%
\begin{align}
\partial_{0}\triangleright(fg)  &  =(\partial_{0}\triangleright f)g+f(\partial
_{0}\triangleright g),\nonumber\\
\partial_{1}\triangleright(fg)  &  =(\partial_{1}\triangleright f)g+(\Lambda
\triangleright f)(\partial_{1}\triangleright g),\\[0.16in]
\hat{\partial}_{0}\,\bar{\triangleright}\,(fg)  &  =(\hat{\partial}_{0}%
\,\bar{\triangleright}\,f)g+f(\hat{\partial}_{0}\,\bar{\triangleright
}\,g),\nonumber\\
\hat{\partial}_{1}\,\bar{\triangleright}\,(fg)  &  =(\hat{\partial}_{1}%
\,\bar{\triangleright}\,f)g+(\Lambda^{-1}\triangleright f)(\hat{\partial}%
_{1}\,\bar{\triangleright}\,g),
\end{align}
and%
\begin{align}
(fg)\triangleleft\hat{\partial}_{0}  &  =f(g\triangleleft\hat{\partial}%
_{0})+(f\triangleleft\hat{\partial}_{0})g,\nonumber\\
(fg)\triangleleft\hat{\partial}_{1}  &  =f(g\triangleleft\hat{\partial}%
_{1})+(f\triangleleft\hat{\partial}_{1})(g\triangleleft\Lambda),\\[0.16in]
(fg)\,\bar{\triangleleft}\,\partial_{0}  &  =f(g\,\bar{\triangleleft
}\,\partial_{0})+(f\,\bar{\triangleleft}\,\partial_{0})g,\nonumber\\
(fg)\,\bar{\triangleleft}\,\partial_{1}  &  =f(g\,\bar{\triangleleft
}\,\partial_{1})+(f\,\bar{\triangleleft}\,\partial_{1})(g\triangleleft
\Lambda^{-1}).
\end{align}
Hitting the above equations with the corresponding integral operator and
rearranging terms, we get%
\begin{align}
(\partial_{0})^{-1}\big |_{x^{0}=\,a}^{b}\triangleright(\partial
_{0}\triangleright f)g  &  =fg\big |_{x^{0}=\,a}^{b}-(\partial_{0}%
)^{-1}\big |_{x^{0}=\,a}^{b}\triangleright f(\partial_{0}\triangleright
g),\nonumber\\
(\partial_{1})^{-1}\big |_{x^{1}=\,a}^{b}\,\triangleright(\partial
_{1}\triangleright f)g  &  =fg\big |_{x^{1}=\,a}^{b}-(\partial_{1}%
)^{-1}\big |_{x^{1}=\,a}^{b}\triangleright(\Lambda\triangleright
f)(\partial_{1}\triangleright g),\\[0.16in]
(\hat{\partial}_{0})^{-1}\big |_{x^{0}=\,a}^{b}\,\bar{\triangleright}%
\,(\hat{\partial}_{0}\,\bar{\triangleright}\,f)g  &  =fg\big |_{x^{0}=\,a}%
^{b}-(\hat{\partial}_{0})^{-1}\big |_{x^{0}=\,a}^{b}\,\bar{\triangleright
}\,f(\hat{\partial}_{0}\,\bar{\triangleright}\,g),\nonumber\\
(\hat{\partial}_{1})^{-1}\big |_{x^{1}=\,a}^{b}\,\bar{\triangleright}%
\,(\hat{\partial}_{1}\,\bar{\triangleright}\,f)g  &  =fg\big |_{x^{1}=\,a}%
^{b}-(\hat{\partial}_{1})^{-1}\big |_{x^{1}=\,a}^{b}\,\bar{\triangleright
}\,(\Lambda^{-1}\triangleright f)(\hat{\partial}_{1}\,\bar{\triangleright
}\,g),
\end{align}
and%
\begin{align}
f(g\triangleleft\hat{\partial}_{0})\triangleleft(\hat{\partial}_{0}%
)^{-1}\big |_{x^{0}=\,a}^{b}  &  =fg\big |_{x^{0}=\,a}^{b}-(f\triangleleft
\hat{\partial}_{0})g\triangleleft(\hat{\partial}_{0})^{-1}\big |_{x^{0}%
=\,a}^{b},\nonumber\\
f(g\triangleleft\hat{\partial}_{1})\triangleleft(\hat{\partial}_{1}%
)^{-1}\big |_{x^{1}=\,a}^{b}  &  =fg\big |_{x^{1}=\,a}^{b}-(f\triangleleft
\hat{\partial}_{1})(g\triangleleft\Lambda)\triangleleft(\hat{\partial}%
_{1})^{-1}\big |_{x^{1}=\,a}^{b},\\[0.16in]
f(g\,\bar{\triangleleft}\,\partial_{0})\,\bar{\triangleleft}\,(\partial
_{0})^{-1}\big |_{x^{0}=\,a}^{b}  &  =fg\big |_{x^{0}=\,a}^{b}-(f\,\bar
{\triangleleft}\,\partial_{0})g\,\bar{\triangleleft}\,(\partial_{0}%
)^{-1}\big |_{x^{0}=\,a}^{b},\nonumber\\
f(g\,\bar{\triangleleft}\,\partial_{1})\,\bar{\triangleleft}\,(\partial
_{1})^{-1}\big |_{x^{1}=\,a}^{b}  &  =fg\big |_{x^{1}=\,a}^{b}-(f\,\bar
{\triangleleft}\,\partial_{1})(g\triangleleft\Lambda^{-1})\,\bar
{\triangleleft}\,(\partial_{1})^{-1}\big |_{x^{1}=\,a}^{b}.
\end{align}

Before we can apply these formulae it remains to write down the explicit form
of the action of the scaling operator. A short glance at the identities in
(\ref{ActLam}) should tell us that%
\begin{align}
\Lambda\triangleright f(x^{i})  &  =f(x^{i})\triangleleft\Lambda^{-1}%
=f(x^{0},qx^{1}),\nonumber\\
\Lambda^{-1}\triangleright f(x^{i})  &  =f(x^{i})\triangleleft\Lambda
=f(x^{0},q^{-1}x^{1}).
\end{align}

We would like to close this subsection by dealing with dual pairings and
q-exponentials. In Ref. \cite{Maj93-5} it was shown that the algebra of
quantum space coordinates and that of the corresponding partial derivatives
are dual to each other. The dual pairings are given by%
\begin{align}
\big \langle f(\partial_{i}),g(x^{j})\big \rangle_{L,\bar{R}}  &
\equiv(f(\partial_{i})\triangleright g(x^{j}))|_{x^{j}=\,0}=(f(\partial
_{i})\,\bar{\triangleleft}\,g(x^{j}))|_{\partial_{i}=\,0},\nonumber\\
\big \langle f(\hat{\partial}_{i}),g(x^{j})\big \rangle_{\bar{L},R}  &
\equiv(f(\hat{\partial}_{i})\,\bar{\triangleright}\,g(x^{j}))|_{x^{j}%
=\,0}=(f(\hat{\partial}_{i})\triangleleft g(x^{j}))|_{\partial_{i}%
=\,0},\\[0.16in]
\big \langle f(x^{i}),g(\partial_{j})\big \rangle_{L,\bar{R}}  &
\equiv(f(x^{i})\,\bar{\triangleleft}\,g(\partial_{j}))|_{x^{i}=\,0}%
=(f(x^{i})\triangleright g(\partial_{j}))|_{\partial_{j}=\,0},\nonumber\\
\big \langle f(x^{i}),g(\hat{\partial}_{j})\big \rangle_{\bar{L},R}  &
\equiv(f(x^{i})\triangleleft g(\hat{\partial}_{j}))|_{x^{i}=\,0}%
=(f(x^{i})\,\bar{\triangleright}\,g(\hat{\partial}_{j}))|_{\partial_{j}=\,0}.
\end{align}
On monomials we get
\begin{align}
\big \langle(\partial_{0})^{n_{0}}(\partial_{1})^{n_{1}},(X^{0})^{m_{0}}%
(X^{1})^{m_{1}}\big \rangle_{L,\bar{R}}  &  =n_{0}![[n_{1}]]_{q}%
!\,\delta^{n_{0},m_{0}}\delta^{n_{1},m_{1}},\nonumber\\
\big \langle(\hat{\partial}_{0})^{n_{0}}(\hat{\partial}_{1})^{n_{1}}%
,(X^{0})^{m_{0}}(X^{1})^{m_{1}}\big \rangle_{\bar{L},R}  &  =n_{0}%
![[n_{1}]]_{q^{-1}}!\,\delta^{n_{0},m_{0}}\delta^{n_{1},m_{1}},
\label{ExpDualAnf}%
\end{align}
and%
\begin{align}
\big \langle(X^{0})^{m_{0}}(X^{1})^{m_{1}},(\partial_{0})^{n_{0}}(\partial
_{1})^{n_{1}}\big \rangle_{L,\bar{R}}  &  =(-1)^{n_{0}+n_{1}}n_{0}%
![[n_{1}]]_{q}!\,\delta^{n_{0},m_{0}}\delta^{n_{1},m_{1}},\nonumber\\
\big \langle(X^{0})^{m_{0}}(X^{1})^{m_{1}},(\hat{\partial}_{0})^{n_{0}}%
(\hat{\partial}_{1})^{n_{1}}\big \rangle_{\bar{L},R}  &  =(-1)^{n_{0}+n_{1}%
}n_{0}![[n_{1}]]_{q^{-1}}!\,\delta^{n_{0},m_{0}}\delta^{n_{1},m_{1}}.
\label{ExplDualEnd}%
\end{align}
These equalities can easily be checked by the identities in
(\ref{ActPartBrai1})-(\ref{ActPartBraid2}).

Now, let us make contact with q-deformed exponentials. From an abstract point
of view an exponential is nothing other than an object whose dualization is
one of the above pairings. In this sense, the exponential is given by the
expression%
\begin{equation}
\exp(x^{i}|\partial_{j})\equiv\sum_{a}e^{a}\otimes f_{a}, \label{ExpAl1N}%
\end{equation}
or%
\begin{equation}
\exp(\partial_{i}|x^{j})\equiv\sum_{a}f_{a}\otimes e^{a}, \label{ExpAl2N}%
\end{equation}
where $\{e_{a}\}$ is a basis in the coordinate algebra and $\{f^{a}\}$ a dual
basis in the algebra of partial derivatives.

If we want to derive explicit formulae for q-deformed exponentials it is our
task to determine a basis of the coordinate algebra being dual to a given one
of the algebra of derivatives. Inserting the elements of the two bases into
the expressions (\ref{ExpAl1N}) and (\ref{ExpAl2N}) will then provide us with
formulae for q-deformed exponentials. It should be obvious that the two bases
being dually paired depend on the choice of the pairing. Thus, each pairing in
(\ref{ExpDualAnf}) and (\ref{ExplDualEnd}) leads to its own q-exponential:%
\begin{align}
&  \big\langle f(\partial_{i}),g(x^{j})\big\rangle_{L,\bar{R}} &  &
\Rightarrow &  &  \exp(x^{i}|\partial_{j})_{\bar{R},L},\nonumber\\
&  \big\langle f(\hat{\partial}_{i}),g(x^{j})\big\rangle_{\bar{L},R} &  &
\Rightarrow &  &  \exp(x^{i}|\hat{\partial}_{j})_{R,\bar{L}},\\[0.16in]
&  \big\langle f(x^{i}),g(\partial_{j})\big\rangle_{L,\bar{R}} &  &
\Rightarrow &  &  \exp(\partial_{i}|x^{j})_{\bar{R},L},\nonumber\\
&  \big\langle f(x^{i}),g(\hat{\partial}_{j})\big\rangle_{\bar{L},R} &  &
\Rightarrow &  &  \exp(\hat{\partial}_{i}|x^{j})_{R,\bar{L}}.
\end{align}
From the results in (\ref{ExpDualAnf}) and (\ref{ExplDualEnd}) we can rather
easily read off two dually paired bases. Proceeding in the above mentioned way
we find%
\begin{align}
\exp(x^{i}|\partial_{j})_{\bar{R},L}  &  =\sum_{n_{0},n_{1}=0}^{\infty}%
\frac{1}{n_{0}![[n_{1}]]_{q}!}(x^{0})^{n_{0}}(x^{1})^{n_{1}}\otimes
(\partial_{0})^{n_{0}}(\partial_{1})^{n_{1}}\nonumber\\
&  =\exp(x^{0}\otimes\partial_{0})\cdot\exp_{q}(x^{1}\otimes\partial
_{1}),\label{Exp1dimAnf}\\[0.16in]
\exp(x^{i}|\hat{\partial}_{j})_{R,\bar{L}}  &  =\sum_{n_{0},n_{1}=0}^{\infty
}\frac{1}{n_{0}![[n_{1}]]_{q^{-1}}!}(x^{0})^{n_{0}}(x^{1})^{n_{1}}\otimes
(\hat{\partial}_{0})^{n_{0}}(\hat{\partial}_{1})^{n_{1}}\nonumber\\
&  =\exp(x^{0}\otimes\hat{\partial}_{0})\cdot\exp_{q^{-1}}(x^{1}\otimes
\hat{\partial}_{1}),
\end{align}
and%
\begin{align}
\exp(\partial_{i}|x^{j})_{\bar{R},L}  &  =\sum_{n_{0},n_{1}=0}^{\infty}%
\frac{1}{n_{0}![[n_{1}]]_{q^{-1}}!}(\partial_{0})^{n_{0}}(\partial_{1}%
)^{n_{1}}\otimes(x^{0})^{n_{0}}(x^{1})^{n_{1}}\nonumber\\
&  =\exp(\partial_{0}\otimes x^{0})\cdot\exp_{q^{-1}}(\partial_{1}\otimes
x^{1}),\\[0.16in]
\exp(\hat{\partial}_{i}|x^{j})_{R,\bar{L}}  &  =\sum_{n_{0},n_{1}=0}^{\infty
}\frac{1}{n_{0}![[n_{1}]]_{q}!}(\hat{\partial}_{0})^{n_{0}}(\hat{\partial}%
_{1})^{n_{1}}\otimes(x^{0})^{n_{0}}(x^{1})^{n_{1}}\nonumber\\
&  =\exp(\hat{\partial}_{0}\otimes x^{0})\cdot\exp_{q}(\hat{\partial}%
_{1}\otimes x^{1}). \label{Exp1dimEnd}%
\end{align}

\subsection{Extended three-dimensional q-deformed Euclidean space}

In this subsection we collect the elements of q-analysis to the extended
three-dimensional q-deformed Euclidean space. Before we can do so, we first
have to answer the question how to perform calculations on an algebra of
non-commutative coordinates - in the following denoted by $\mathcal{A}_{q}$.
This can be accomplished by a kind of pullback that transforms operations on
the non-commutative coordinate algebra to those on a commutative one. For this
to become more clear, one should realize that the non-commutative coordinate
algebra\ we are dealing with satisfies the \textit{Poincar\'{e}-Birkhoff-Witt
property}. It tells us that the dimension of a subspace of homogeneous
polynomials has to be the same as for commuting coordinates. This property is
the deeper reason why monomials of a given normal ordering constitute a basis
of the non-commutative algebra $\mathcal{A}_{q}$. Due to this fact we can
establish a vector space isomorphism between $\mathcal{A}_{q}$ and a
commutative algebra $\mathcal{A}$ generated by ordinary coordinates
$x^{1},x^{2},\ldots,x^{n}$:
\begin{align}
\mathcal{W}  &  :\mathcal{A}\longrightarrow\mathcal{A}_{q},\nonumber\\
\mathcal{W}((x^{1})^{i_{1}}\ldots(x^{n})^{i_{n}})  &  \equiv(X^{1})^{i_{1}%
}\ldots(X^{n})^{i_{n}}. \label{AlgIso}%
\end{align}

This vector space isomorphism can even be extended to an algebra isomorphism
by introducing a non-commutative product in $\mathcal{A}$, the so-called
\textit{star product} \cite{BFF78, Moy49, MSSW00}. This product is defined via
the relation
\begin{equation}
\mathcal{W}(f\circledast g)=\mathcal{W}(f)\cdot\mathcal{W}(g),
\end{equation}
being tantamount to%
\begin{equation}
f\circledast g\equiv\mathcal{W}^{-1}\left(  \mathcal{W}\left(  f\right)
\cdot\mathcal{W}\left(  g\right)  \right)  ,
\end{equation}
where $f$ and $g$ are formal power series in $\mathcal{A}$.

In the case of the extended three-dimensional q-deformed Euclidean space it is
convenient to work with the normal orderings%
\begin{equation}
\mathcal{W}\left(  (x^{0})^{n_{0}}(x^{+})^{n_{+}}(x^{3})^{n_{3}}(x^{-}%
)^{n_{-}}\right)  =(X^{0})^{n_{0}}(X^{+})^{n_{+}}(X^{3})^{n_{3}}(X^{-}%
)^{n_{-}}, \label{sternid1}%
\end{equation}
and%
\begin{equation}
\widetilde{\mathcal{W}}\left(  (x^{0})^{n_{0}}(x^{+})^{n_{+}}(x^{3})^{n_{3}%
}(x^{-})^{n_{-}}\right)  =(X^{0})^{n_{0}}(X^{-})^{n_{-}}(X^{3})^{n_{3}}%
(X^{+})^{n_{+}}.
\end{equation}
The star product corresponding to the first choice takes the form \cite{WW01}%
\begin{align}
f(x^{i})\circledast g(x^{j})=\,  &  \sum_{k=0}^{\infty}\lambda^{k}\frac
{(x^{3})^{2k}}{[[k]]_{q^{4}}!}q^{2(\hat{n}_{x^{3}}\hat{n}_{y^{+}}+\,\hat
{n}_{x^{-}}\hat{n}_{y^{3}})}\nonumber\label{sternformel}\\
\,  &  \times\,\left.  (D_{q^{4}}^{-})^{k}f(x^{i})\cdot(D_{q^{4}}^{+}%
)^{k}g(y^{j})\right\vert _{y\rightarrow x},
\end{align}
and likewise for the second choice,%
\begin{align}
\tilde{f}(x^{i})\circledast\tilde{g}(x^{j})=\,  &  \sum_{k=0}^{\infty}\left(
-\lambda\right)  ^{k}\frac{(x^{3})^{2k}}{[[k]]_{q^{-4}}!}q^{-2(\hat{n}_{x^{3}%
}\hat{n}_{y^{+}}+\,\hat{n}_{x^{-}}\hat{n}_{y^{3}})}\nonumber\\
\,  &  \times\,\left.  (D_{q^{-4}}^{+})^{k}\tilde{f}(x^{i})\cdot(D_{q^{-4}%
}^{-})^{k}\tilde{g}(y^{j})\right\vert _{y\rightarrow x}.
\end{align}
Notice that the tilde on top of the symbols for the functions shall remind us
of the fact that the star product refers to the algebra homomorphism
$\widetilde{\mathcal{W}}$. Extending the three-dimensional q-deformed
Euclidean space by a time element does not really change the operator
expressions for the star product. This observation is a consequence of the
fact that the time element is central in the algebra of quantum space coordinates.

In very much the same way as was done for the braided line we can calculate
actions of partial derivatives on normally ordered monomials by applying the
commutation relations in (\ref{Lei3dimExp1})-(\ref{Lei3dimExp2}). By means of
the algebra isomorphisms (\ref{AlgIso}) these actions carry over to
commutative functions, i.e. we have
\begin{align}
\partial_{i}\triangleright\mathcal{W}(f)  &  =\mathcal{W}(\partial
_{i}\triangleright f),\quad f\in\mathcal{A}\text{,}\nonumber\\
\mathcal{W}(f)\triangleleft\partial_{i}  &  =\mathcal{W}(f\triangleleft
\partial_{i}),
\end{align}
or%
\begin{align}
\partial_{i}\triangleright f  &  \equiv\mathcal{W}^{-1}\left(  \partial
_{i}\triangleright\mathcal{W}(f)\right)  ,\nonumber\\
f\triangleleft\partial_{i}  &  \equiv\mathcal{W}^{-1}\left(  \mathcal{W}%
(f)\triangleleft\partial_{i}\right)  .
\end{align}

In the work of Ref. \cite{BW01} we derived operator representations
of\ q-deformed partial derivatives by applying these ideas. The results for
the q-deformed three-dimensional Euclidean space can easily be modified to
include the time element $X^{0}$ and the corresponding partial derivative
$\partial_{0}$. In doing so we get%
\begin{align}
\partial_{0}\triangleright f  &  =\frac{\partial}{\partial x^{0}}f,\nonumber\\
\partial_{+}\triangleright f  &  =D_{q^{4}}^{+}f,\nonumber\\
\partial_{3}\triangleright f  &  =D_{q^{2}}^{3}f(q^{2}x^{+}),\nonumber\\
\partial_{-}\triangleright f  &  =D_{q^{4}}^{-}f(q^{2}x^{3})+\lambda
x^{+}(D_{q^{2}}^{3})^{2}f. \label{DarstAbl}%
\end{align}

The expressions for the other types of actions of partial derivatives follow
from the above formulae by applying the substitutions%
\begin{align}
\partial_{i}\triangleright f  &  \overset{{%
\genfrac{}{}{0pt}{}{\pm}{q}%
}{%
\genfrac{}{}{0pt}{}{\rightarrow}{\rightarrow}%
}{%
\genfrac{}{}{0pt}{}{\mp}{1/q}%
}}{\longleftrightarrow}\hat{\partial}_{\overline{i}}\,\bar{\triangleright
}\,\tilde{f},\nonumber\\
f\,\bar{\triangleleft}\,\partial_{i}  &  \overset{{%
\genfrac{}{}{0pt}{}{\pm}{q}%
}{%
\genfrac{}{}{0pt}{}{\rightarrow}{\rightarrow}%
}{%
\genfrac{}{}{0pt}{}{\mp}{1/q}%
}}{\longleftrightarrow}\tilde{f}\triangleleft\hat{\partial}_{\overline{i}},
\label{TransRule1}%
\end{align}
and%
\begin{align}
\partial_{i}\triangleright f  &  \overset{+\leftrightarrow-}%
{\longleftrightarrow}f\,\bar{\triangleleft}\,\partial_{\overline{i}%
},\nonumber\\
\hat{\partial}_{i}\,\bar{\triangleright}\,\tilde{f}  &  \overset
{+\leftrightarrow-}{\longleftrightarrow}\tilde{f}\triangleleft\hat{\partial
}_{\overline{i}}, \label{TransRule2}%
\end{align}
where the symbols $\overset{{%
\genfrac{}{}{0pt}{}{\pm}{q}%
}{%
\genfrac{}{}{0pt}{}{\rightarrow}{\rightarrow}%
}{%
\genfrac{}{}{0pt}{}{\mp}{1/q}%
}}{\longleftrightarrow}$ and $\overset{+\leftrightarrow-}{\longleftrightarrow
}$ respectively denote transitions via the substitutions%
\begin{equation}
D_{q^{a}}^{\pm}\rightarrow D_{q^{-a}}^{\mp},\quad x^{\pm}\rightarrow x^{\mp
},\quad q\rightarrow q^{-1},
\end{equation}
and%
\begin{equation}
D_{q^{a}}^{\pm}\rightarrow D_{q^{a}}^{\mp},\quad x^{\pm}\rightarrow x^{\mp}.
\end{equation}
Notice that in (\ref{TransRule1}) and (\ref{TransRule2}) we introduced a
conjugate index with%
\begin{equation}
\overline{(+,3,-,0)}=(-,3,+,0).
\end{equation}

Now, we come to integrals for the extended three-dimensional q-deformed
Euclidean space. For this reason we enhance the algebra of partial derivatives
by introducing inverse elements. The additional relations then read%
\begin{align}
(\partial_{i})^{-1}\partial_{i}  &  =\partial_{i}(\partial_{i})^{-1}%
=1,\nonumber\\
(\partial_{0})^{-1}\partial_{i}  &  =\partial_{i}(\partial_{0})^{-1}%
,\nonumber\\
(\partial_{i})^{-1}\partial_{0}  &  =\partial_{0}(\partial_{i})^{-1},\quad
i\in\{+,3,-,0\},\nonumber\\
(\partial_{3})^{-1}\partial_{\pm}  &  =q^{\pm2}\partial_{\pm}(\partial
_{3})^{-1},\nonumber\\
(\partial_{\pm})^{-1}\partial_{3}  &  =q^{\mp2}\partial_{3}(\partial_{\pm
})^{-1},\nonumber\\
(\partial_{+})^{-1}\partial_{-}  &  =\partial_{-}(\partial_{+})^{-1}%
-q^{-4}\lambda(\partial_{3})^{2}(\partial_{+})^{-2},\nonumber\\
\partial_{+}(\partial_{-})^{-1}  &  =(\partial_{-})^{-1}\partial_{+}%
-q^{-4}\lambda(\partial_{-})^{-2}(\partial_{3})^{2}.
\end{align}

As a next step we would like to find representations for the inverse partial
derivatives. From a short glance at (\ref{DarstAbl}) it should become obvious
that
\begin{align}
(\partial_{0})^{-1}\big |_{x^{0}=\,a}^{b}\triangleright f  &  =\int_{a}%
^{b}dx^{0}\,f,\nonumber\\
(\partial_{+})^{-1}\big |_{x^{+}=\,a}^{b}\triangleright f  &  =(D_{q^{4}}%
^{+})^{-1}\big |_{x^{+}=\,a}^{b}f,\nonumber\\
(\partial_{3})^{-1}\big |_{x^{3}=\,a}^{b}\triangleright f  &  =(D_{q^{3}}%
^{3})^{-1}\big |_{x^{3}=\,a}^{b}f(q^{-2}x^{+}).
\end{align}

It remains to derive the representation corresponding to\ $(\partial_{-}%
)^{-1}$. To this end, the representation of $\partial_{-}$ is divided up into
a classical part and corrections vanishing in the undeformed limit
$q\rightarrow1$, i.e.%
\begin{equation}
\partial_{-}\triangleright f=(\partial_{-})_{\text{cl}}f+(\partial
_{-})_{\text{cor}}f,
\end{equation}
where
\begin{equation}
(\partial_{-})_{\text{cl}}f=D_{q^{4}}^{-}f(q^{2}x^{3}),\quad(\partial
_{-})_{\text{cor}}f=\lambda x^{+}(D_{q^{2}}^{3})^{2}f.
\end{equation}
Then we can proceed as follows:%
\begin{align}
(\partial_{-})^{-1}\triangleright f  &  =\frac{1}{(\partial_{-})_{\text{cl}%
}+(\partial_{-})_{\text{cor}}}f=\frac{1}{(\partial_{-})_{\text{cl}}\left(
1+(\partial_{-})_{\text{cl}}^{-1}(\partial_{-})_{\text{cor}}\right)
}f\nonumber\\
&  =\frac{1}{1+(\partial_{-})_{\text{cl}}^{-1}(\partial_{-})_{\text{cor}}%
}\cdot\frac{1}{(\partial_{-})_{\text{cl}}}f\nonumber\\
&  =\sum_{k=0}^{\infty}\left(  -1\right)  ^{k}\left[  ((\partial
_{-})_{\text{cl}}^{-1}(\partial_{-})_{\text{cor}}\right]  ^{k}((\partial
_{-})_{\text{cl}})^{-1}f\nonumber\label{IntegralE3N}\\
&  =\sum_{k=0}^{\infty}q^{2k(k+1)}(-\lambda x^{+})^{k}(D_{q^{2}}^{3}%
)^{2k}(D_{q^{4}}^{-})^{-(k+1)}f(q^{-2(k+1)}x^{3}).
\end{align}

In complete analogy to the correspondences in (\ref{TransRule1}) and
(\ref{TransRule2}) the other types of representations follow from the above
formulae by applying the transformation rules%
\begin{align}
(\partial_{i})^{-1}\triangleright f  &  \overset{{%
\genfrac{}{}{0pt}{}{\pm}{q}%
}{%
\genfrac{}{}{0pt}{}{\rightarrow}{\rightarrow}%
}{%
\genfrac{}{}{0pt}{}{\mp}{1/q}%
}}{\longleftrightarrow}(\hat{\partial}_{\overline{i}})^{-1}\,\bar
{\triangleright}\,\tilde{f},\nonumber\\
f\,\bar{\triangleleft}\,(\partial_{i})^{-1}  &  \overset{{%
\genfrac{}{}{0pt}{}{\pm}{q}%
}{%
\genfrac{}{}{0pt}{}{\rightarrow}{\rightarrow}%
}{%
\genfrac{}{}{0pt}{}{\mp}{1/q}%
}}{\longleftrightarrow}\tilde{f}\triangleleft(\hat{\partial}_{\overline{i}%
})^{-1},
\end{align}
and%
\begin{align}
(\partial_{i})^{-1}\triangleright f  &  \overset{+\leftrightarrow
-}{\longleftrightarrow}f\,\bar{\triangleleft}\,(\partial_{\overline{i}}%
)^{-1},\nonumber\\
(\hat{\partial}_{i})^{-1}\,\bar{\triangleright}\,\tilde{f}  &  \overset
{+\leftrightarrow-}{\longleftrightarrow}\tilde{f}\triangleleft(\hat{\partial
}_{\overline{i}})^{-1}.
\end{align}

Next, we would like to concern ourselves with q-translations on the
three-dimensional q-deformed Euclidean space. We already mentioned that the
Hopf structures on quantum space coordinates imply their translations. For the
coproducts on quantum space coordinates we have%
\begin{align}
\Delta_{L}(X^{0})=\,  &  X^{0}\otimes1+1\otimes X^{0},\nonumber\\
\Delta_{L}(X^{-})=\,  &  X^{-}\otimes1+\Lambda^{-1/2}\tau^{-1/2}\otimes
X^{-},\nonumber\\
\Delta_{L}(X^{3})=\,  &  X^{3}\otimes1+\Lambda^{-1/2}\otimes X^{3}%
+\lambda\lambda_{+}\Lambda^{-1/2}L^{+}\otimes X^{-},\nonumber\\
\Delta_{L}(X^{+})=\,  &  X^{-}\otimes1+\Lambda^{-1/2}\tau^{-1/2}\otimes
X^{+}+q\lambda\lambda_{+}\Lambda^{-1/2}\tau^{1/2}L^{+}\otimes X^{3}\nonumber\\
\,  &  +\,q^{-2}\lambda^{2}\lambda_{+}\Lambda^{-1/2}\tau^{1/2}(L^{+}%
)^{2}\otimes X^{-},
\end{align}
and
\begin{align}
\Delta_{\bar{L}}(X^{0})=\,  &  X^{0}\otimes1+1\otimes X^{0},\nonumber\\
\Delta_{\bar{L}}(X^{+})=\,  &  X^{+}\otimes1+\Lambda^{1/2}\tau^{-1/2}\otimes
X^{+},\nonumber\\
\Delta_{\bar{L}}(X^{3})=\,  &  X^{3}\otimes1+\Lambda^{1/2}\otimes
X^{3}+\lambda\lambda_{+}\Lambda^{1/2}L^{-}\otimes X^{+},\nonumber\\
\Delta_{\bar{L}}(X^{-})=\,  &  X^{-}\otimes1+\Lambda^{1/2}\tau^{1/2}\otimes
X^{-}+q^{-1}\lambda\lambda_{+}\Lambda^{1/2}\tau^{1/2}L^{-}\otimes
X^{3}\nonumber\\
\,  &  +\,q^{-2}\lambda^{2}\lambda_{+}\Lambda^{1/2}\tau^{1/2}(L^{-}%
)^{2}\otimes X^{+}, \label{CoproConN}%
\end{align}
where\ $L^{+},$ $L^{-},$ and $\tau$ denote generators of $U_{q}(su_{2}),$
while $\Lambda$ plays the role of a scaling operator with ($A=\{+,3,-\}$),%
\begin{gather}
\Lambda X^{0}=X^{0}\Lambda,\quad\Lambda X^{A}=q^{4}X^{A}\Lambda,\nonumber\\
\Lambda\partial_{0}=\partial_{0}\Lambda,\quad\Lambda\partial_{A}%
=q^{-4}\partial_{A}\Lambda.
\end{gather}
The corresponding antipodes take the form%
\begin{align}
S_{L}(X^{0})=\,  &  -X^{0},\nonumber\\
S_{L}(X^{-})=\,  &  -\Lambda^{1/2}\tau^{1/2}X^{-},\nonumber\\
S_{L}(X^{3})=\,  &  -\Lambda^{1/2}X^{3}+q^{2}\lambda\lambda_{+}\Lambda
^{1/2}\tau^{1/2}L^{+}X^{-},\nonumber\\
S_{L}(X^{+})=\,  &  -\Lambda^{1/2}\tau^{-1/2}X^{-}+q\lambda\lambda_{+}%
\Lambda^{1/2}L^{+}X^{3}\nonumber\\
\,  &  -\,q^{4}\lambda^{2}\lambda_{+}\Lambda^{1/2}\tau^{1/2}(L^{+})^{2}X^{-},
\end{align}
and%
\begin{align}
S_{\bar{L}}(X^{0})=  &  -X^{0},\nonumber\\
S_{\bar{L}}(X^{+})=  &  -\Lambda^{-1/2}\tau^{1/2}X^{+},\nonumber\\
S_{\bar{L}}(X^{3})=  &  -\Lambda^{-1/2}X^{3}+q^{-2}\lambda\lambda_{+}%
\Lambda^{-1/2}\tau^{1/2}L^{-}X^{+},\nonumber\\
S_{\bar{L}}(X^{-})=  &  -\Lambda^{-1/2}\tau^{-1/2}X^{-}+q^{-1}\lambda
\lambda_{+}\Lambda^{-1/2}L^{-}X^{3}\nonumber\\
&  -q^{-4}\lambda^{2}\lambda_{+}\Lambda^{-1/2}\tau^{1/2}(L^{-})^{2}X^{+}.
\end{align}

We see that coproduct and antipode become rather simple on the time element.
This observation is a direct consequence of the fact that the time element is
completely decoupled from position space. For the same reason the Hopf
structures on the subspace spanned by the coordinates $X^{+},$ $X^{3},$ and
$X^{-}$ are identical to those already presented in the work of Ref.
\cite{Wac04}.

It is not very difficult to modify the reasonings in Ref. \cite{Wac04} in a
way that they take account of the existence of the time element $X^{0}$. In
this manner, we can show that the above relations imply
\begin{align}
&  f(x^{i}\oplus_{\bar{L}}y^{j})=\sum_{k_{0}=0}^{n_{0}}\sum_{k_{+}=0}^{n_{+}%
}\sum_{k_{3}=0}^{n_{3}}\sum_{k_{-}=0}^{n_{-}}\sum_{l=0}^{k_{3}}(q\lambda
\lambda_{+})^{l}\nonumber\\
&  \qquad\times\frac{(x^{0})^{k_{0}}(x^{+})^{k_{+}}(x^{3})^{k_{3}-l}%
(x^{-})^{k_{-}+\,l}(y^{+})^{l}}{k_{0}![[2i]]_{q^{2}}!![[k_{+}]]_{q^{4}%
}![[k_{3}-l]]_{q^{2}}![[k_{-}]]_{q^{4}}!}\nonumber\\
&  \qquad\times\,\Big ((D_{q^{4}}^{+})^{k_{+}}(D_{q^{2}}^{3})^{k_{3}%
+l}(D_{q^{4}}^{-})^{k_{-}}\Big (\frac{\partial}{\partial x^{0}}\Big )^{k_{0}%
}f\Big )(q^{2(k_{3}-l)}y^{+},q^{2k_{-}}y^{3}). \label{KomCoprDreiN}%
\end{align}
and
\begin{align}
&  \hat{U}(f(\ominus_{\bar{L}}\,x^{i}))=\sum_{k=0}^{\infty}(q^{-1}%
\lambda\lambda_{+})^{k}q^{4k^{2}}\,\frac{(x^{+}x^{-})^{k}}{[[2k]]_{q^{2}}%
!!}\nonumber\\
&  \qquad\times(D_{q^{2}}^{3})^{2k}\,q^{2(\hat{n}_{+}^{2}+\,\hat{n}_{-}%
^{2})+\hat{n}_{3}(2\hat{n}_{+}+\,2\hat{n}_{-}+\,\,\hat{n}_{3}-1)}%
\,f(-x^{0},-x^{+},-q^{-2k}x^{3},-x^{-}), \label{KomAntDreiN}%
\end{align}
where
\begin{equation}
\lbrack\lbrack2k]]_{q^{2}}!!=[[2k]]_{q^{2}}[[2(k-1)]]_{q^{2}}\ldots
\lbrack\lbrack2]]_{q^{2}}.
\end{equation}

The operator $\hat{U}$ in Eq. (\ref{KomAntDreiN}) transforms a function of
normal ordering $X^{+}X^{3}X^{-}$ into another function representing the same
element but now for reversed ordering. Its explicit form\ was presented in the
work of Ref.\ \cite{BW01}. The expressions corresponding to the other Hopf
structures are obtained from the formulae in (\ref{KomCoprDreiN}) and
(\ref{KomAntDreiN}) most easily by means of the transitions%
\begin{equation}
f(x^{i}\oplus_{\bar{L}}y^{j})\overset{{%
\genfrac{}{}{0pt}{}{\pm}{q}%
}{%
\genfrac{}{}{0pt}{}{\rightarrow}{\rightarrow}%
}{%
\genfrac{}{}{0pt}{}{\mp}{1/q}%
}}{\longleftrightarrow}\tilde{f}(x^{i}\,\oplus_{L}\,y^{j}),
\end{equation}
and
\begin{equation}
\hat{U}(f(\ominus_{\bar{L}}\,x^{i}))\overset{{%
\genfrac{}{}{0pt}{}{\pm}{q}%
}{%
\genfrac{}{}{0pt}{}{\rightarrow}{\rightarrow}%
}{%
\genfrac{}{}{0pt}{}{\mp}{1/q}%
}}{\longleftrightarrow}\hat{U}^{-1}(\tilde{f}(\ominus_{L}\,x^{i})).
\end{equation}
The tilde again reminds us of the fact that the function refers to reversed
normal ordering.

Next, we would like to say a few words about braided products on the extended
q-deformed Euclidean space in three dimensions. As we know, braided products
describe how elements of different quantum spaces commute. In this sense they
are an essential ingredient to formulate multiplication on tensor products of
quantum spaces. The entries $\mathcal{L}_{j}^{i}$ of the so-called L-matrix
determine the braiding of the quantum space coordinates $X^{i},$
$i\in\{0,+,3,-\}$ (if not stated otherwise summation over repeated indices is
to be understood):%
\begin{equation}
X^{i}\odot_{L}w=(\mathcal{L}_{j}^{i}\triangleright w)\otimes X^{j}.
\end{equation}
The explicit form of the L-matrix can be read off from the coproduct on
coordinates, since it holds%
\begin{equation}
\Delta_{L}(X^{i})=X^{i}\otimes1+\mathcal{L}_{j}^{i}\otimes X^{j}.
\end{equation}
In very much the same way we have%
\begin{equation}
X^{i}\odot_{\bar{L}}w=(\mathcal{\bar{L}}_{j}^{i}\triangleright w)\otimes
X^{j},
\end{equation}
and%
\begin{equation}
\Delta_{\bar{L}}(X^{i})=X^{i}\otimes1+\mathcal{\bar{L}}_{j}^{i}\otimes X^{j}.
\end{equation}

These considerations are consistent with the observation that the time
coordinate $X^{0}$ shows trivial braiding. In Ref. \cite{Wac05} we presented
operator expressions that realize braided products for the three-dimensional
q-deformed Euclidean space on a commutative coordinate algebra. Due to the
trivial braiding of the time coordinate these expressions carry over to the
extended three-dimensional q-deformed Euclidean space without any changes.

Last but not least, we come to dual pairings and q-exponentials. We already
recalled their definition in Sec.\thinspace\ref{QAnBraid}. With the results of
Ref. \cite{Wac03} it is not very difficult to show that%
\begin{align}
&  \big \langle(\partial_{0})^{n_{0}}(\partial_{-})^{n_{-}}(\partial
_{3})^{n_{3}}(\partial_{+})^{n_{-}},(X^{0})^{m_{0}}(X^{+})^{m_{+}}%
(X^{3})^{m_{3}}(X^{-})^{m_{-}}\big \rangle_{L,\bar{R}}=\nonumber\\
&  \qquad=\delta_{m_{-},n_{-}}\delta_{m_{3},n_{3}}\delta_{m_{+},n_{+}}%
\delta_{m_{0},n_{0}}m_{0}!\,[[m_{+}]]_{q^{4}}!\,[[m_{3}]]_{q^{2}}%
!\,[[m_{-}]]_{q^{4}}!,\\[0.16in]
&  \big \langle(\hat{\partial}_{0})^{n_{0}}(\hat{\partial}_{+})^{n_{+}}%
(\hat{\partial}_{3})^{n_{3}}(\hat{\partial}_{-})^{n_{-}},(X^{0})^{m_{0}}%
(X^{-})^{m_{-}}(X^{3})^{m_{3}}(X^{+})^{m_{+}}\big \rangle_{\bar{L}%
,R}=\nonumber\\
&  \qquad=\delta_{m_{-},n_{-}}\delta_{m_{3},n_{3}}\delta_{m_{+},n_{+}}%
\delta_{m_{0},n_{0}}m_{0}!\,[[m_{+}]]_{q^{-4}}!\,[[m_{3}]]_{q^{-2}}%
!\,[[m_{-}]]_{q^{-4}}!,
\end{align}
and%
\begin{align}
&  \big \langle(X^{0})^{m_{0}}(X^{+})^{m_{+}}(X^{3})^{m_{3}}(X^{-})^{m_{-}%
},(\partial_{0})^{n_{0}}(\partial_{-})^{n_{-}}(\partial_{3})^{n_{3}}%
(\partial_{+})^{n_{+}}\big \rangle_{L,\bar{R}}=\nonumber\\
&  \qquad=(-1)^{n_{0}+n_{+}+n_{3}+n_{-}}\delta_{m_{-},n_{-}}\delta
_{m_{3},n_{3}}\delta_{m_{+},n_{+}}\delta_{m_{0},n_{0}}\nonumber\\
&  \qquad\hspace{0.16in}\times m_{0}!\,[[m_{+}]]_{q^{4}}!\,[[m_{3}]]_{q^{2}%
}!\,[[m_{-}]]_{q^{4}}!,\\[0.16in]
&  \big \langle(X^{0})^{m_{0}}(X^{-})^{m_{-}}(X^{3})^{m_{3}}(X^{+})^{m_{+}%
},(\hat{\partial}_{0})^{n_{0}}(\hat{\partial}_{+})^{n_{+}}(\hat{\partial}%
_{3})^{n_{3}}(\hat{\partial}_{-})^{n_{-}}\big \rangle_{L,\bar{R}}=\nonumber\\
&  \qquad=(-1)^{n_{0}+n_{+}+n_{3}+n_{-}}\delta_{m_{-},n_{-}}\delta
_{m_{3},n_{3}}\delta_{m_{+},n_{+}}\delta_{m_{0},n_{0}}\nonumber\\
&  \qquad\hspace{0.16in}\times m_{0}!\,[[m_{+}]]_{q^{-4}}!\,[[m_{3}]]_{q^{-2}%
}!\,[[m_{-}]]_{q^{-4}}!.
\end{align}
From these pairings we can read off the explicit form of q-exponentials for
the extended three-dimensional q-deformed Euclidean space:%
\begin{align}
&  \exp(x^{i}|\partial_{j})_{\bar{R},L}=\nonumber\\
&  \qquad=\sum_{\underline{n}=0}^{\infty}\frac{(x^{0})^{n_{0}}(x^{+})^{n_{+}%
}(x^{3})^{n_{3}}(x^{-})^{n_{-}}\otimes(\partial_{0})^{n_{0}}(\partial
_{-})^{n_{-}}(\partial_{3})^{n_{3}}(\partial_{+})^{n_{+}}}{n_{0}%
!\,[[n_{+}]]_{q^{4}}!\,[[n_{3}]]_{q^{2}}!\,[[n_{-}]]_{q^{4}}!}%
,\label{Exp3dimAnf}\\[0.1in]
&  \exp(x^{i}|\hat{\partial}_{j})_{R,\bar{L}}=\nonumber\\
&  \qquad=\sum_{\underline{n}=0}^{\infty}\frac{(x^{0})^{n_{0}}(x^{-})^{n_{-}%
}(x^{3})^{n_{3}}(x^{+})^{n_{+}}\otimes(\hat{\partial}_{0})^{n_{0}}%
(\hat{\partial}_{+})^{n_{+}}(\hat{\partial}_{3})^{n_{3}}(\hat{\partial}%
_{-})^{n_{-}}}{n_{0}!\,[[n_{+}]]_{q^{-4}}!\,[[n_{3}]]_{q^{-2}}!\,[[n_{-}%
]]_{q^{-4}}!}.
\end{align}
The expressions for the other exponentials follow from these formulae by
applying the transformations%
\begin{align}
&  \exp(x^{i}|\partial_{j})_{\bar{R},L}{}\overset{+\leftrightarrow
-}{\longleftrightarrow}\exp(\partial_{i}|x^{j})_{\bar{R},L},\nonumber\\
&  \exp(x^{i}|\hat{\partial}_{j})_{R,\bar{L}}{}\overset{+\leftrightarrow
-}{\longleftrightarrow}\exp(\hat{\partial}_{i}|x^{j})_{R,\bar{L}%
},\label{Exp3dimEnd}%
\end{align}
where the symbol $\overset{+\leftrightarrow-}{\longleftrightarrow}$ denotes a
transition between the two expressions via one of the following substitutions:%
\begin{align}
\text{a)}\quad X^{i} &  \leftrightarrow-\partial_{i},\quad\partial
_{i}\leftrightarrow X^{i},\nonumber\\
\text{b)}\quad X^{i} &  \leftrightarrow-\hat{\partial}_{i},\quad\hat{\partial
}_{i}\leftrightarrow X^{i}.
\end{align}

\section{Time evolution operator\label{SecTimEvo}}

In this section we discuss the question how wave functions on the quantum
spaces under consideration change in time. First of all, we recall that
translations on quantum spaces are generated by q-exponentials \cite{qAn,
Maj95, Maj93-5, Wac04, SW04}. This observation leads us to q-deformed Taylor
rules which take the form \cite{qAn}%
\begin{align}
\exp(x^{i}\oplus_{\bar{L}}(\ominus_{\bar{L}}\,y^{j})|\partial_{k})_{\bar{R}%
,L}\overset{\partial|y}{\triangleright}g(y^{l})  &  =g(x^{i}),\nonumber\\
\exp(x^{i}\oplus_{L}(\ominus_{L}\,y^{j})|\hat{\partial}_{k})_{R,\bar{L}%
}\,\overset{\partial|y}{\bar{\triangleright}}\,g(y^{k})  &  =g(x^{i}),
\label{q-TayRec}%
\end{align}
and%
\begin{align}
g(y^{l})\,\overset{y|\partial}{\bar{\triangleleft}}\,\exp(\partial
_{k}|(\ominus_{R}\,y^{j})\oplus_{R}x^{i})_{\bar{R},L}  &  =g(x^{i}%
),\nonumber\\
g(y^{l})\overset{y|\partial}{\triangleleft}\exp(\hat{\partial}_{k}%
|(\ominus_{\bar{R}}\,y^{j})\oplus_{\bar{R}}x^{i})_{R,\bar{L}}  &  =g(x^{i}).
\label{q-TayRecN}%
\end{align}
For a correct understanding of these expressions see also Ref. \cite{qAn,
Wac04}.

If the q-deformed Taylor rules shall describe translations in time, only, they
have to be modified as follows:%
\begin{align}
\big [\exp(x^{i}\oplus_{\bar{L}}(\ominus_{\bar{L}}\,y^{j})|\partial_{k}%
)_{\bar{R},L}\overset{\partial|y}{\triangleright}g(y^{l})\big ]_{x^{A}%
=\,y^{A}}  &  =g(y^{i})\big |_{y^{0}=\,x^{0}},\nonumber\\
\big [\exp(x^{i}\oplus_{L}(\ominus_{L}\,y^{j})|\hat{\partial}_{k})_{R,\bar{L}%
}\,\overset{\partial|y}{\bar{\triangleright}}\,g(y^{l})\big ]_{x^{A}=\,y^{A}}
&  =g(y^{i})\big |_{y^{0}=\,x^{0}},
\end{align}
and%
\begin{align}
\big [g(y^{l})\,\overset{y|\partial}{\bar{\triangleleft}}\,\exp(\partial
_{k}|(\ominus_{R}\,y^{j})\oplus_{R}x^{i})_{\bar{R},L}\big ]_{x^{A}=\,y^{A}}
&  =g(y^{i})\big |_{y^{0}=\,x^{0}},\nonumber\\
\big [g(y^{l})\overset{y|\partial}{\triangleleft}\exp(\hat{\partial}%
_{k}|(\ominus_{\bar{R}}\,y^{j})\oplus_{\bar{R}}x^{i})_{R,\bar{L}}%
\big ]_{x^{A}=\,y^{A}}  &  =g(y^{i})\big |_{y^{0}=\,x^{0}},
\end{align}
where $A$ represents the indices $(+,3,-)$. In the above expressions we first
perform a general translation and then require that the space coordinates of
the translated function take on the same values as the original function.
Since space and time are completely decoupled from each other the above
formulae simplify to
\begin{align}
\big [\exp(x^{0}\otimes\partial_{0})\overset{\partial|y}{\triangleright
}g(y^{i})\big ]_{y^{0}=\,0}  &  =g(y^{i})\big |_{y^{0}=\,x^{0}},\nonumber\\
\big [\exp(x^{0}\otimes\hat{\partial}_{0})\overset{\partial|y}{\triangleright
}g(y^{i})\big ]_{y^{0}=\,0}  &  =g(y^{i})\big |_{y^{0}=\,x^{0}},
\end{align}
and%
\begin{align}
\big [g(y^{i})\,\overset{y|\partial}{\bar{\triangleleft}}\,\exp(-\partial
_{0}\otimes x^{0})\big ]_{y^{0}=\,0}  &  =g(y^{i})\big |_{y^{0}=\,x^{0}%
},\nonumber\\
\big [g(y^{i})\overset{y|\partial}{\triangleleft}\exp(-\hat{\partial}%
_{0}\otimes x^{0})\big ]_{y^{0}=\,0}  &  =g(y^{i})\big |_{y^{0}=\,x^{0}}.
\end{align}

In quantum mechanics the set of values a wave function takes on in space at a
certain time completely determines the behavior of that wave function at all
later times. This requires that time derivatives acting on wave functions can
be substituted by a linear operator i$^{-1}H$ that acts on space coordinates,
only, and has the same algebraic properties as the time derivative
$\partial_{0}$.

In this manner, it should be clear that for the time evolution operator we
have%
\begin{align}
\phi(x^{A},t)  &  =\mathcal{U}_{L}(t,t^{\prime}=0)\overset{x}{\triangleright
}\phi(x^{A},t^{\prime}=0)\nonumber\\
&  =\mathcal{U}_{\bar{L}}(t,t^{\prime}=0)\overset{x}{\triangleright}\phi
(x^{A},t^{\prime}=0)\nonumber\\
&  =\phi(x^{A},t^{\prime}=0)\overset{x}{\triangleleft}\mathcal{U}%
_{R}(t,t^{\prime}=0)\nonumber\\
&  =\phi(x^{A},t^{\prime}=0)\overset{x}{\triangleleft}\mathcal{U}_{\bar{R}%
}(t,t^{\prime}=0), \label{TimEvoId1}%
\end{align}
where%
\begin{align}
\mathcal{U}_{L}(t,t^{\prime}  &  =0)=\mathcal{U}_{\bar{L}}(t,t^{\prime
}=0)\equiv\exp(-t\otimes\text{i}H),\\[0.08in]
\mathcal{U}_{R}(t,t^{\prime}  &  =0)=\mathcal{U}_{\bar{R}}(t,t^{\prime
}=0)\equiv\exp(\text{i}H\otimes t).
\end{align}
We see that the time evolution operator is of the same form as in the
undeformed case. In the remainder of this section we collect basic properties
of the time evolution operators. This is mainly done for the purpose of
providing consistent notation.

First of all, we are seeking operators $\mathcal{U}_{\gamma}^{-1}(t,t^{\prime
}=0),$ $\gamma\in\{L,\bar{L},R,\bar{R}\},$ with\
\begin{align}
&  \mathcal{U}_{\gamma}(t,t^{\prime}=0)\,\mathcal{U}_{\gamma}^{-1}%
(t,t^{\prime}=0)=1,\nonumber\\
&  \mathcal{U}_{\gamma}^{-1}(t,t^{\prime}=0)\,\mathcal{U}_{\gamma}%
(t,t^{\prime}=0)=1.
\end{align}
One readily checks that%
\begin{align}
&  \mathcal{U}_{\alpha}^{-1}(t,t^{\prime}=0)\equiv\mathcal{U}_{\alpha
}(-t,t^{\prime}=0)=\exp(t\otimes\text{i}H),\\[0.08in]
&  \mathcal{U}_{\beta}^{-1}(t,t^{\prime}=0)\equiv\mathcal{U}_{\beta
}(-t,t^{\prime}=0)=\exp(-\text{i}H\otimes t),
\end{align}
where $\alpha\in\{L,\bar{L}\}$ and $\beta\in\{R,\bar{R}\}.$ As a direct
consequence of these identities we have%
\begin{align}
\phi(x^{A},t^{\prime}=0)  &  =\,\mathcal{U}_{\alpha}^{-1}(t,t^{\prime
}=0)\triangleright\phi(x^{A},t)\nonumber\\
&  =\,\phi(x^{A},t)\triangleleft\mathcal{U}_{\beta}^{-1}(t,t^{\prime}=0).
\label{TimEvoId2N}%
\end{align}
The operators $\mathcal{U}_{\gamma}^{-1}(t,t^{\prime}=0)$ describe particles
traversing backwards in time, since we have%
\begin{align}
\phi(x^{A},-t)  &  =\,\mathcal{U}_{\alpha}^{-1}(t,t^{\prime}=0)\triangleright
\phi(x^{A},t^{\prime}=0)\nonumber\\
&  =\,\phi(x^{A},t^{\prime}=0)\triangleleft\mathcal{U}_{\beta}^{-1}%
(t,t^{\prime}=0).
\end{align}

Now, we are in a position to generalize the time evolution operators by
\begin{align}
&  \mathcal{U}_{\alpha}(t,t^{\prime})\equiv\mathcal{U}_{\alpha}(t,t^{\prime
\prime}=0)\,\mathcal{U}_{\alpha}^{-1}(t^{\prime},t^{\prime\prime}%
=0)=\exp(-(t-t^{\prime})\otimes\text{i}H),\label{GenTimEvo1}\\[0.08in]
&  \mathcal{U}_{\beta}(t,t^{\prime})\equiv\mathcal{U}_{\beta}^{-1}(t^{\prime
},t^{\prime\prime}=0)\,\mathcal{U}_{\beta}(t,t^{\prime\prime}=0)=\exp
(-\text{i}H\otimes(t^{\prime}-t)). \label{GenTimEvo2}%
\end{align}
The new operators tell us how wave functions change under a time displacement
$t^{\prime}\rightarrow t$:%
\begin{equation}
\phi(x^{A},t)=\mathcal{U}_{\alpha}(t,t^{\prime})\triangleright\phi
(x^{A},t^{\prime})=\phi(x^{A},t^{\prime})\triangleleft\mathcal{U}_{\beta
}(t,t^{\prime}). \label{TimDisp}%
\end{equation}
To prove these equalities one can apply the identities in (\ref{TimEvoId1})
and (\ref{TimEvoId2N}). An essential feature of the time evolution operators
in (\ref{GenTimEvo1}) and (\ref{GenTimEvo2}) is the composition property,%
\begin{align}
\mathcal{U}_{\alpha}(t,t^{\prime})  &  =\mathcal{U}_{\alpha}(t,0)\,\mathcal{U}%
_{\alpha}^{-1}(t^{\prime},0)\nonumber\\
&  =\mathcal{U}_{\alpha}(t,0)\,\mathcal{U}_{\alpha}^{-1}(t^{\prime\prime
},0)\,\mathcal{U}_{\alpha}(t^{\prime\prime},0)\,\mathcal{U}_{\alpha}%
^{-1}(t^{\prime},0)\nonumber\\
&  =\mathcal{U}_{\alpha}(t,t^{\prime\prime})\,\mathcal{U}_{\alpha}%
(t^{\prime\prime},t^{\prime}),
\end{align}
and%
\begin{align}
\mathcal{U}_{\beta}(t,t^{\prime})  &  =\mathcal{U}_{\beta}^{-1}(t^{\prime
},0)\,\mathcal{U}_{\beta}(t,0)\nonumber\\
&  =\mathcal{U}_{\beta}^{-1}(t^{\prime},0)\,\mathcal{U}_{\beta}(t^{\prime
\prime},0)\,\mathcal{U}_{\beta}^{-1}(t^{\prime\prime},0)\,\mathcal{U}_{\beta
}(t,0)\nonumber\\
&  =\mathcal{U}_{\beta}(t^{\prime\prime},t^{\prime})\,\mathcal{U}_{\beta
}(t,t^{\prime\prime}).
\end{align}

Next, we would like to consider operators $\mathcal{U}_{\gamma}^{-1}%
(t,t^{\prime})$ being subject to%
\begin{equation}
\mathcal{U}_{\gamma}(t,t^{\prime})\,\mathcal{U}_{\gamma}^{-1}(t,t^{\prime
})=\mathcal{U}_{\gamma}^{-1}(t,t^{\prime})\,\mathcal{U}_{\gamma}(t,t^{\prime
})=1.
\end{equation}
They are given by%
\begin{align}
\mathcal{U}_{\alpha}^{-1}(t,t^{\prime})  &  \equiv\mathcal{U}_{\alpha
}(t^{\prime},t^{\prime\prime}=0)\,\mathcal{U}_{\alpha}^{-1}(t,t^{\prime\prime
}=0)=\mathcal{U}_{\alpha}(t^{\prime},t)\nonumber\\
&  =\mathcal{U}_{\alpha}(-t,-t^{\prime})=\exp((t-t^{\prime})\otimes
\text{i}H),\\[0.16in]
\mathcal{U}_{\beta}^{-1}(t,t^{\prime})  &  \equiv\mathcal{U}_{\beta}%
^{-1}(t,t^{\prime\prime}=0)\,\mathcal{U}_{\beta}(t^{\prime},t^{\prime\prime
}=0)=\mathcal{U}_{\beta}(t^{\prime},t)\nonumber\\
&  =\mathcal{U}_{\beta}(-t,-t^{\prime})=\exp(\text{i}H\otimes(t^{\prime}-t)).
\end{align}
These operators reverse the time displacement $t^{\prime}\rightarrow t$:
\begin{equation}
\phi(x^{A},t^{\prime})=\mathcal{U}_{\alpha}^{-1}(t,t^{\prime})\triangleright
\phi(x^{A},t)=\phi(x^{A},t)\triangleleft\mathcal{U}_{\beta}^{-1}(t,t^{\prime
}).
\end{equation}

Last but not least we would like to mention some simplifications. Let us
recall that the time coordinate shows trivial braiding. Thus, the tensor
products in the expressions on the right-hand side of (\ref{GenTimEvo1}) and
(\ref{GenTimEvo2}) can be omitted. Concretely, we can make the identifications%
\begin{equation}
\exp(t\otimes\text{i}H)=\exp(\text{i}H\otimes t)=\exp(\text{i}Ht),
\end{equation}
and%
\begin{align}
\mathcal{U}(t,t^{\prime})  &  =\mathcal{U}_{\alpha}(t,t^{\prime}%
)=\mathcal{U}_{\beta}(-t,-t^{\prime})\nonumber\\
&  =\mathcal{U}_{\alpha}^{-1}(-t,-t^{\prime})=\mathcal{U}_{\beta}%
^{-1}(t,t^{\prime}),
\end{align}
where
\begin{equation}
\mathcal{U}(t,t^{\prime})=\exp(-\text{i}H(t-t^{\prime})).
\end{equation}
It is obvious that the operator $\mathcal{U}(t,t^{\prime})$ becomes unitary,
if the Hamiltonian $H$ is assumed to be Hermitian:%
\begin{equation}
\mathcal{U}^{-1}(t,t^{\prime})=\mathcal{U}(-t,-t^{\prime})=\mathcal{U}%
(t^{\prime},t)=\mathcal{U}^{\dag}(t,t^{\prime}).
\end{equation}

\section{Schr\"{o}dinger and Heisenberg picture\label{SHPic}}

In the last section we found that the time evolution operator on the quantum
spaces under consideration is of the same form as its undeformed counterpart.
For this reason we should be able to introduce the Heisenberg and the
Schr\"{o}dinger picture on our quantum spaces along the same line of
reasonings as in the undeformed case (see for example Ref. \cite{Sak94}).

To begin with we derive differential equations for the time evolution
operators. We have
\begin{align}
\partial_{0}\overset{t}{\triangleright}\mathcal{U}_{L}(t,t^{\prime})  &
=\partial_{0}\overset{t}{\triangleright}\mathcal{U}_{L}(t,0)\,\mathcal{U}%
_{L}^{-1}(t^{\prime},0)\nonumber\\
&  =\partial_{0}\overset{t}{\triangleright}\exp(-t\otimes\text{i}%
H)\,\mathcal{U}_{L}^{-1}(t^{\prime},0)\nonumber\\
&  =\exp(-t\otimes\text{i}H)\,(-\text{i}H)\,\mathcal{U}_{L}^{-1}(t^{\prime
},0)\nonumber\\
&  =-\text{i}H\exp(-t\otimes\text{i}H)\,\mathcal{U}_{L}^{-1}(t^{\prime
},0)\nonumber\\
&  =-\text{i}H\,\mathcal{U}_{L}(t,0)\,\mathcal{U}_{L}^{-1}(t^{\prime
},0)\nonumber\\
&  =-\text{i}H\,\mathcal{U}_{L}(t,t^{\prime}),
\end{align}
and, likewise,
\begin{align}
\mathcal{U}_{R}(t,t^{\prime})\overset{t}{\triangleleft}\hat{\partial}_{0}  &
=\mathcal{U}_{R}^{-1}(t^{\prime},0)\,\mathcal{U}_{R}(t,0)\overset
{t}{\triangleleft}\hat{\partial}_{0}\nonumber\\
&  =\mathcal{U}_{R}^{-1}(t^{\prime},0)\,\exp(\text{i}H\otimes t)\overset
{t}{\triangleleft}\hat{\partial}_{0}\nonumber\\
&  =\mathcal{U}_{R}^{-1}(t^{\prime},0)\,(-\text{i}H)\,\exp(\text{i}H\otimes
t)\nonumber\\
&  =\mathcal{U}_{R}^{-1}(t^{\prime},0)\,\exp(\text{i}H\otimes t)\,(-\text{i}%
H)\nonumber\\
&  =\mathcal{U}_{R}^{-1}(t^{\prime},0)\,\mathcal{U}_{R}(t,0)\,(-\text{i}%
H)\nonumber\\
&  =\mathcal{U}_{R}(t,t^{\prime})\,(-\text{i}H).
\end{align}
In this manner, we find\ that%
\begin{align}
\text{i}\partial_{0}\overset{t}{\triangleright}\mathcal{U}_{L}(t,t^{\prime})
&  =H\,\mathcal{U}_{L}(t,t^{\prime}),\nonumber\\
\text{i}\hat{\partial}_{0}\,\overset{t}{\bar{\triangleright}}\,\mathcal{U}%
_{\bar{L}}(t,t^{\prime})  &  =H\,\mathcal{U}_{\bar{L}}(t,t^{\prime}),
\label{SchoEq1}%
\end{align}
and%
\begin{align}
\mathcal{U}_{R}(t,t^{\prime})\overset{t}{\triangleleft}(\text{i}\hat{\partial
}_{0})  &  =\mathcal{U}_{R}(t,t^{\prime})H,\nonumber\\
\mathcal{U}_{\bar{R}}(t,t^{\prime})\,\overset{t}{\bar{\triangleleft}%
}\,(\text{i}\partial_{0})  &  =\mathcal{U}_{\bar{R}}(t,t^{\prime})H.
\label{SchroEq2}%
\end{align}

The above equations, which are often referred to as Schr\"{o}dinger equations
of the time evolution operator, correspond to different geometries. However,
from the considerations so far one can conclude that the equations in
(\ref{SchoEq1}) and (\ref{SchroEq2}) are not really different from each other.
Thus, the reader may think that such a distinction is unnecessary. But this is
not the case, since the realization of the Hamiltonian often depends on the
choice for the geometry.

It should also be mentioned that the differential equations in (\ref{SchoEq1})
and (\ref{SchroEq2}) are equivalent to the integral equations%
\begin{align}
\mathcal{U}_{\alpha}(t,t^{\prime})  &  =1-\text{i}\int_{t^{\prime}}%
^{t}dt^{\prime\prime}\,H\,\mathcal{U}_{\alpha}(t^{\prime\prime},t^{\prime
}),\nonumber\\
\mathcal{U}_{\beta}(t,t^{\prime})  &  =1+\text{i}\int_{t^{\prime}}%
^{t}dt^{\prime\prime}\,\mathcal{U}_{\beta}(t^{\prime\prime},t^{\prime})H,
\end{align}
if we require%
\begin{equation}
\mathcal{U}_{\gamma}(t,t)=1.
\end{equation}
Formal solutions are given by%
\begin{align}
\mathcal{U}_{\alpha}(t,t^{\prime})  &  =1+\sum_{n=1}^{\infty}\text{i}^{-n}%
\int_{t^{\prime}}^{t}dt_{1}\int_{t^{\prime}}^{t_{1}}dt_{2}\ldots
\int_{t^{\prime}}^{t_{n-1}}dt_{n}\,H(t_{1})H(t_{2})\ldots H(t_{n}),\nonumber\\
\mathcal{U}_{\beta}(t,t^{\prime})  &  =1+\sum_{n=1}^{\infty}\text{i}^{n}%
\int_{t^{\prime}}^{t}dt_{1}\int_{t^{\prime}}^{t_{1}}dt_{2}\ldots
\int_{t^{\prime}}^{t_{n-1}}dt_{n}\,H(t_{n})H(t_{n-1})\ldots H(t_{1}).
\end{align}

Let us recall that in the Schr\"{o}dinger picture wave functions vary with
time, while observables like $X^{i}$ and $P^{i}$ are fixed in time. We obtain
the equations of motion in the Schr\"{o}dinger picture by combining
(\ref{TimDisp}) with the equations in (\ref{SchoEq1}) and (\ref{SchroEq2}).
Proceeding in this manner yields%
\begin{align}
\text{i}\partial_{0}\overset{t}{\triangleright}\phi(x^{A},t)  &
=\text{i}\partial_{0}\overset{t}{\triangleright}\mathcal{U}_{L}(t,t^{\prime
})\triangleright\phi(x^{A},t^{\prime})\nonumber\\
&  =H\,\mathcal{U}_{L}(t,t^{\prime})\triangleright\phi(x^{A},t^{\prime
})=H\overset{x}{\triangleright}\phi(x^{A},t),\\[0.1in]
\text{i}\hat{\partial}_{0}\,\overset{t}{\bar{\triangleright}}\,\phi(t,x^{A})
&  =\text{i}\hat{\partial}_{0}\,\overset{t}{\bar{\triangleright}}%
\,\mathcal{U}_{\bar{L}}(t,t^{\prime})\triangleright\phi(x^{A},t^{\prime
})\nonumber\\
&  =H\,\mathcal{U}_{\bar{L}}(t,t^{\prime})\triangleright\phi(x^{A},t^{\prime
})=H\,\overset{x}{\bar{\triangleright}}\,\phi(x^{A},t),
\end{align}
and
\begin{align}
\phi(x^{A},t)\overset{t}{\triangleleft}(\text{i}\hat{\partial}_{0})  &
=\phi(x^{A},t^{\prime})\triangleleft\,\mathcal{U}_{R}(t,t^{\prime})\overset
{t}{\triangleleft}\hat{\partial}_{0}\nonumber\\
&  =\phi(x^{A},t^{\prime})\triangleleft\,\mathcal{U}_{R}(t,t^{\prime}%
)H=\phi(x^{A},t)\overset{x}{\triangleleft}H,\\[0.1in]
\phi(x^{A},t)\,\overset{t}{\bar{\triangleleft}}\,(\text{i}\partial_{0})  &
=\phi(x^{A},t^{\prime})\triangleleft\,\mathcal{U}_{\bar{R}}(t,t^{\prime
})\,\overset{t}{\bar{\triangleleft}}\,(\text{i}\partial_{0})\nonumber\\
&  =\phi(x^{A},t^{\prime})\triangleleft\,\mathcal{U}_{\bar{R}}(t,t^{\prime
})H=\phi(x^{A},t)\,\overset{x}{\bar{\triangleleft}}\,H.
\end{align}

Next, we would like to discuss the implications of these equations on the time
dependence of transition amplitudes and expectation values. To this end, we
first introduce sesquilinear forms on the quantum spaces under consideration.
In analogy to the undeformed case they can be defined by \cite{WQK1}%
\begin{align}
\big \langle f,g\big \rangle_{\gamma}  &  \equiv\int_{-\infty}^{+\infty
}d_{\gamma}^{n}x\,\overline{f(x^{A},t)}\overset{t,x}{\circledast}%
g(x^{B},t),\nonumber\\
\big \langle f,g\big \rangle_{\gamma}^{\prime}  &  \equiv\int_{-\infty
}^{+\infty}d_{\gamma}^{n}x\,f(x^{A},t)\overset{t,x}{\circledast}%
\overline{g(x^{B},t)}, \label{PraSes}%
\end{align}
where again $\gamma\in\{L,\bar{L},R,\bar{R}\}.$ For the integrals over the
whole space we have to insert the expressions \cite{Wac02, Wac04, qAn}

\begin{itemize}
\item[(i)] (braided line)%
\begin{align}
\int_{-\infty}^{+\infty}d_{L}x\,f(x^{A},t)  &  =(D_{q}^{1})^{-1}%
\big |_{-\infty}^{\infty}\,f\nonumber\\
&  =-\int_{-\infty}^{+\infty}d_{\bar{R}}x\,f(x^{A},t),\\[0.1in]
\int_{-\infty}^{+\infty}d_{\bar{L}}x\,f(x^{A},t)  &  =(D_{q^{-1}}^{1}%
)^{-1}\big |_{-\infty}^{\infty}\,f\nonumber\\
&  =-\int_{-\infty}^{+\infty}d_{R}x\,f(x^{A},t),
\end{align}

\item[(ii)] (q-deformed Euclidean space)%
\begin{align}
\int_{-\infty}^{+\infty}d_{L}^{3}x\,f(x^{A},t)  &  =\frac{q^{-6}}{4}(D_{q^{2}%
}^{+})^{-1}\big |_{-\infty}^{\infty}(D_{q^{2}}^{3})^{-1}\big |_{-\infty
}^{\infty}(D_{q^{2}}^{-})^{-1}\big |_{-\infty}^{\infty}\,f\nonumber\\
&  =-\int_{-\infty}^{+\infty}d_{\bar{R}}^{3}x\,f(t,x^{A}),\\[0.1in]
\int_{-\infty}^{+\infty}d_{\bar{L}}^{3}x\,f(x^{A},t)  &  =\frac{q^{6}}%
{4}(D_{q^{-2}}^{-})^{-1}\big |_{-\infty}^{\infty}(D_{q^{-2}}^{3}%
)^{-1}\big |_{-\infty}^{\infty}(D_{q^{-2}}^{+})^{-1}\big |_{-\infty}^{\infty
}\,f\nonumber\\
&  =-\int_{-\infty}^{+\infty}d_{R}^{3}x\,f(x^{A},t),
\end{align}
\textbf{ }However, there is one difficulty we have to overcome here. The
conjugation properties of q-deformed integrals are responsible for the fact
that the sesquilinear forms in (\ref{PraSes}) are not symmetrical \cite{qAn,
WQK1}. To circumvent this problem one can take the sesquilinear forms%
\begin{align}
\big \langle f,g\big \rangle_{1}  &  \equiv\frac{\text{i}^{n}}{2}%
\big (\big \langle f,g\big \rangle_{L}+\big \langle f,g\big \rangle_{\bar{R}%
}\big ),\nonumber\\
\big \langle f,g\big \rangle_{2}  &  \equiv\frac{\text{i}^{n}}{2}%
\big (\big \langle f,g\big \rangle_{\bar{L}}+\big \langle f,g\big \rangle_{R}%
\big ),\label{SymSes1}\\[0.16in]
\big \langle f,g\big \rangle_{1}^{\prime}  &  \equiv\frac{\text{i}^{n}}%
{2}\big (\big \langle f,g\big \rangle_{L}^{\prime}%
+\big \langle f,g\big \rangle_{\bar{R}}^{\prime}\big ),\nonumber\\
\big \langle f,g\big \rangle_{2}^{\prime}  &  \equiv\frac{\text{i}^{n}}%
{2}\big (\big \langle f,g\big \rangle_{\bar{L}}^{\prime}%
+\big \langle f,g\big \rangle_{R}^{\prime}\big ). \label{SymSes2}%
\end{align}

\end{itemize}

Clearly, all information on the time development of a sesquilinear form is
contained in the time dependence of its arguments. Normally, the time
evolution operators are unitary, so sesquilinear forms of two wave functions
should not vary with time. In complete analogy to the undeformed case we have
$(i=1,2)$%
\begin{align}
\big \langle\phi,\psi\big \rangle_{i}  &  \equiv\int_{-\infty}^{+\infty}%
d_{i}^{n}x\,\overline{\phi(x^{A},t)}\overset{t,x}{\circledast}\psi
(x^{B},t)\nonumber\\
&  =\int_{-\infty}^{+\infty}d_{i}^{n}x\,\overline{\mathcal{U}(t,t^{\prime
})\triangleright\phi(x^{A},t^{\prime})}\overset{t,x}{\circledast}%
(\mathcal{U}(t,t^{\prime})\triangleright\psi(x^{B},t^{\prime}))\nonumber\\
&  =\int_{-\infty}^{+\infty}d_{i}^{n}x\,(\,\overline{\phi(x^{A},t^{\prime}%
)}\triangleleft\mathcal{U}^{\dag}(t,t^{\prime}))\overset{t,x}{\circledast
}(\mathcal{U}(t,t^{\prime})\triangleright\psi(x^{B},t^{\prime}))\nonumber\\
&  =\int_{-\infty}^{+\infty}d_{i}^{n}x\,\overline{\phi(x^{A},t^{\prime}%
)}\overset{t^{\prime}\!,x}{\circledast}(\mathcal{U}^{-1}(t,t^{\prime
})\,\mathcal{U}(t,t^{\prime})\triangleright\psi(x^{B},t^{\prime}))\nonumber\\
&  =\int_{-\infty}^{+\infty}d_{i}^{n}x\,\overline{\phi(x^{A},t^{\prime}%
)}\overset{t^{\prime}\!,x}{\circledast}\psi(x^{B},t^{\prime})\nonumber\\
&  =\int_{-\infty}^{+\infty}d_{i}^{n}x\,\overline{\phi(x^{A},t^{\prime}%
=0)}\overset{x}{\circledast}\psi(x^{B},t^{\prime}=0)\nonumber\\
&  =\big \langle\phi,\psi\big \rangle_{i}\big |_{t=0.} \label{CalTimDevSes}%
\end{align}
where, for brevity, we introduced%
\begin{align}
\int_{-\infty}^{+\infty}d_{1}^{n}x  &  \equiv\frac{\text{i}^{n}}{2}%
\Big (\int_{-\infty}^{+\infty}d_{L}^{n}x+\int_{-\infty}^{+\infty}d_{\bar{R}%
}^{n}x\Big ),\nonumber\\
\int_{-\infty}^{+\infty}d_{2}^{n}x  &  \equiv\frac{\text{i}^{n}}{2}%
\Big (\int_{-\infty}^{+\infty}d_{\bar{L}}^{n}x+\int_{-\infty}^{+\infty}%
d_{R}^{n}x\Big ). \label{SubInt}%
\end{align}
Similar arguments lead us to
\begin{align}
\big \langle\phi,\psi\big \rangle_{i}^{\prime}  &  \equiv\int_{-\infty
}^{+\infty}d_{i}^{n}x\,\phi(x^{A},t)\overset{t,x}{\circledast}\overline
{\psi(x^{B},t)}\nonumber\\
&  =\int_{-\infty}^{+\infty}d_{i}^{n}x\,\phi(x^{A},t^{\prime})\overset
{t^{\prime}\!,x}{\circledast}\overline{\psi(x^{B},t^{\prime})}\nonumber\\
&  =\int_{-\infty}^{+\infty}d_{i}^{n}x\,\phi(x^{A},t^{\prime}=0)\overset
{x}{\circledast}\overline{\psi(x^{B},t^{\prime}=0)}\nonumber\\
&  =\big \langle\phi,\psi\big \rangle_{i}^{\prime}\big |_{t=0.}%
\end{align}
Wee see that on the quantum spaces under consideration wave functions keep
their normalization, i.e. the equalities%
\begin{equation}
\big \langle\phi,\phi\big \rangle_{i,x}=1, \label{NorBed1}%
\end{equation}
or%
\begin{equation}
\big \langle\phi,\phi\big \rangle_{i,x}^{\prime}=1, \label{NorBed2}%
\end{equation}
remain unchanged as time goes by.

Next, we turn attention to matrix elements of observables and examine their
time development. With the same reasonings already applied in
(\ref{CalTimDevSes}) we obtain%
\begin{align}
\big \langle\phi,\hat{O}\triangleright\psi\big \rangle_{i,x}  &
=\int_{-\infty}^{\infty}d_{i}^{n}x\,\overline{\phi(x^{A},t)}\overset
{t,x}{\circledast}(\hat{O}\triangleright\psi(x^{B},t))\nonumber\\
&  =\int_{-\infty}^{\infty}d_{i}^{n}x\,\overline{\mathcal{U}(t,t^{\prime
})\triangleright\phi(x^{A},t^{\prime})}\overset{t,x}{\circledast}(\hat
{O}\triangleright(\mathcal{U}(t,t^{\prime})\triangleright\psi(x^{B},t^{\prime
})))\nonumber\\
&  =\int_{-\infty}^{\infty}d_{i}^{n}x\,\overline{\phi(x^{A},t^{\prime}%
)}\overset{t^{\prime}\!,x}{\circledast}(\mathcal{U}^{-1}(t,t^{\prime}%
)\,\hat{O}\,\mathcal{U}(t,t^{\prime})\triangleright\psi(x^{B},t^{\prime
}))\nonumber\\
&  =\int_{-\infty}^{\infty}d_{i}^{n}x\,\overline{\phi(x^{A},0)}\overset
{x}{\circledast}(\mathcal{U}^{-1}(t,0)\,\hat{O}\,\mathcal{U}%
(t,0)\triangleright\psi(x^{B},0)).
\end{align}
Repeating the same steps for the sesquilinear forms with apostrophe we get%
\begin{align}
\big \langle\phi\triangleleft\hat{O}^{\prime},\psi\big \rangle_{i,x}^{\prime}
&  =\int_{-\infty}^{\infty}d_{i}^{n}x\,(\phi(x^{A},t^{\prime})\triangleleft
\mathcal{U}(t,t^{\prime})\,\hat{O}^{\prime}\,\mathcal{U}^{-1}(t,t^{\prime
}))\overset{t^{\prime},x}{\circledast}\overline{\psi(x^{B},t^{\prime}%
)}\nonumber\\
&  =\int_{-\infty}^{\infty}d_{i}^{n}x\,(\phi(x^{A},0)\triangleleft
\mathcal{U}(t,0)\,\hat{O}^{\prime}\,\mathcal{U}^{-1}(t,0))\overset
{x}{\circledast}\overline{\psi(x^{B},0)}.
\end{align}

The above reasonings show us that the Heisenberg picture can indeed be
introduced in very much the same way as is done in the undeformed case, i.e.
we define the Heisenberg picture observable by%
\begin{equation}
\hat{O}_{H}\equiv\mathcal{U}^{-1}(t,0)\,\hat{O}\,\mathcal{U}(t,0),\qquad
\hat{O}_{H}^{\prime}\equiv\mathcal{U}(t,0)\,\hat{O}^{\prime}\,\mathcal{U}%
^{-1}(t,0),
\end{equation}
while the corresponding wave functions are independent from time:%
\begin{equation}
\phi_{H}(x^{A})\equiv\phi(x^{A},t=0).
\end{equation}
It should be obvious that this convention leads to the same matrix elements
and expectation values as in the Schr\"{o}dinger picture.

In the Heisenberg picture time evolution is assigned to observables and not to
wave functions. Thus, the equations of motion do not concern wave functions
but observables. Realizing that the time derivatives on the quantum spaces
under considerations coincide with those on commutative spaces we regain the
well-known Heisenberg equations of motion, i.e.%
\begin{align}
\frac{d\hat{O}_{H}}{dt}  &  =\frac{\partial\mathcal{U}^{-1}(t,0)}{\partial
t}\,\hat{O}\,\mathcal{U}(t,0)+\mathcal{U}^{-1}(t,0)\,\hat{O}\,\frac
{\partial\mathcal{U}(t,0)}{\partial t}\nonumber\\
&  =\text{i}H\,\mathcal{U}^{-1}(t,0)\,\hat{O}\,\mathcal{U}(t,0)-\,\mathcal{U}%
^{-1}(t,0)\,\hat{O}\,\mathcal{U}(t,0)\,\text{i}H\nonumber\\
&  =\text{i}[H,\hat{O}_{H}],\\[0.16in]
\frac{d\hat{O}_{H}^{\prime}}{dt}  &  =\frac{\partial\mathcal{U}(t,0)}{\partial
t}\,\hat{O}^{\prime}\,\mathcal{U}^{-1}(t,0)+\mathcal{U}(t,0)\,\hat{O}^{\prime
}\,\frac{\partial\mathcal{U}^{-1}(t,0)}{\partial t}\nonumber\\
&  =-\text{i}H\,\mathcal{U}(t,0)\,\hat{O}^{\prime}\,\mathcal{U}^{-1}%
(t,0)+\,\mathcal{U}(t,0)\,\hat{O}^{\prime}\,\mathcal{U}^{-1}(t,0)\,\text{i}%
H\nonumber\\
&  =\text{i}[\hat{O}_{H}^{\prime},H],
\end{align}
where we assumed the Hamiltonian to be time-independent.

\section{Conclusion\label{SecCon}}

Let us end with some comments on what we have done so far. In this article we
enhanced the algebras of braided line and q-deformed three-dimensional
Euclidean space by adding a time element. This was done in a way being
consistent with the existing algebraic framework. We were then able to apply
our reasonings about constructing q-deformed analogs of classical analysis. We
saw that the time element is completely decoupled from space coordinates and
behaves like a commutative variable. In doing so, we arrived at mathematical
structures in which space is discretized while time is still continuous. The
clear distinction between space and time made it easy to develop the basics of
a q-deformed analog of non-relativistic Schr\"{o}dinger theory. Fortunately,
we could apply the same reasonings as in the undeformed case, to which our
results tend in the limit $q\rightarrow1$.

Especially, we found that the time evolution operators are of the same general
form as their undeformed counterpart, i.e. they can again be obtained by
exponentiation of a Hamiltonian. The Schr\"{o}dinger and the Heisenberg
picture could be developed in a rather straightforward way and apart from the
fact that we have different q-geometries we could regain the well-known
equations of motion, i.e. the Schr\"{o}dinger and the Heisenberg equations. In
this manner, we laid the foundations for discretized versions of
non-relativistic quantum mechanics that do not lack space-time symmetries.

Based on the reasonings of part I we will continue this program in part II of
our paper. In this respect, let us point out that compared to other quantum
spaces, like the q-deformed Minkowski space,\ extended braided line and
extended three-dimensional q-deformed Euclidean space provide a rather simple
arena for studying the implications of q-deformation on quantum mechanics and
quantum field theory.

Last but not least, we would like to say a few words about q-deformed
superanalysis on the braided line, since this subject has not been treated up
to now. To this end we have to consider the antisymmetrized space determined
by relation (\ref{DiffBrai}). (However, we are mainly interested in the
subspace that is spanned by the q-deformed Grassmann variable $\theta^{1}$
subject to $(\theta^{1})^{2}=0.)$ In the work of Refs. \cite{MSW04, SW04} it
was described how to construct superanalysis on q-deformed quantum spaces. It
is rather easy to apply these ideas to the antisymmetrized braided line. In
what follows we give a short review of the results of this undertaking.

If we require for the differential $d\equiv d\theta^{i}(\partial_{\theta}%
)_{i}$ to hold%
\begin{align}
d^{2}  &  =0,\nonumber\\
d(fg)  &  =(df)g-f(dg),
\end{align}
the Leibniz rules for the two differential calculi on the\ antisymmetrized
braided line become%
\begin{align}
(\partial_{\theta})_{i}\theta^{j}  &  =\delta_{i}^{j}-\hat{R}_{il}%
^{jk}\,\theta^{l}(\partial_{\theta})_{k},\nonumber\\
(\hat{\partial}_{\theta})_{i}\theta^{j}  &  =\delta_{i}^{j}-(\hat{R}%
^{-1})_{il}^{jk}\,\theta^{l}(\hat{\partial}_{\theta})_{k},
\end{align}
where%
\begin{equation}
(\hat{\partial}_{\theta})_{0}=-(\partial_{\theta})_{0},\quad(\hat{\partial
}_{\theta})_{1}=-q(\partial_{\theta})_{1}.
\end{equation}
Especially, we have%
\begin{align}
(\partial_{\theta})_{1}\theta^{1}  &  =1-q\theta^{1}(\partial_{\theta}%
)_{1},\nonumber\\
(\hat{\partial}_{\theta})_{1}\theta^{1}  &  =1-q^{-1}\theta^{1}(\hat{\partial
}_{\theta})_{1}.
\end{align}

For supernumbers of the form%
\begin{equation}
f(\theta^{1})=f^{\prime}+f_{1}\theta^{1},\quad f^{\prime},f_{1}\in\mathbb{C},
\end{equation}
the actions of antisymmetric partial derivatives take the form%
\begin{equation}
(\partial_{\theta})_{1}\triangleright f(\theta^{1})=(\hat{\partial}_{\theta
})_{1}\,\bar{\triangleright}\,f(\theta^{1})=f_{1},
\end{equation}
and%
\begin{equation}
f(\theta^{1})\,\bar{\triangleleft}\,(\partial_{\theta})_{1}=f(\theta
^{1})\triangleleft(\hat{\partial}_{\theta})_{1}=-f_{1}.
\end{equation}

In complete analogy to the undeformed case integration and differentiation are
the same on q-deformed antisymmetrized spaces :%
\begin{align}
\int d_{L}\theta^{1}\,f(\theta^{1})  &  =(\partial_{\theta})_{1}\triangleright
f(\theta^{1})=f_{1},\nonumber\\
\int d_{\bar{L}}\theta^{1}\,f(\theta^{1})  &  =(\hat{\partial}_{\theta}%
)_{1}\,\bar{\triangleright}\,f(\theta^{1})=f_{1},\\[0.1in]
\int d_{R}\theta^{1}\,f(\theta^{1})  &  =f(\theta^{1})\triangleleft
(\hat{\partial}_{\theta})_{1}=-f_{1},\nonumber\\
\int d_{\bar{R}}\theta^{1}\,f(\theta^{1})  &  =f(\theta^{1})\,\bar
{\triangleleft}\,(\partial_{\theta})_{1}=-f_{1.}%
\end{align}

Following the ideas in Ref. \cite{SW04} translations of q-deformed Grassmann
variables are determined by their Hopf structures, for which we have%
\begin{align}
\Delta_{L}(\theta^{i})  &  =\theta^{i}\otimes1+\tilde{\Lambda}^{-1}%
\otimes\theta^{i},\nonumber\\
S_{L}(\theta^{i})  &  =-\tilde{\Lambda}^{-1}\theta^{i},\nonumber\\
\epsilon_{L}(\theta^{i})  &  =0, \label{HopAntBra1}%
\end{align}
and
\begin{align}
\Delta_{\bar{L}}(\theta^{i})  &  =\theta^{i}\otimes1+\tilde{\Lambda}%
\otimes\theta^{i},\nonumber\\
S_{\bar{L}}(\theta^{i})  &  =-\tilde{\Lambda}\theta^{i},\nonumber\\
\epsilon_{\bar{L}}(\theta^{i})  &  =0, \label{HopAntBra2}%
\end{align}
where the unitary scaling operator $\tilde{\Lambda}$ has to fulfill%
\begin{equation}
\tilde{\Lambda}\theta^{i}=-q^{\delta_{i1}}\theta^{i}\tilde{\Lambda}%
,\qquad\tilde{\Lambda}(\partial_{\theta})_{i}=-q^{-\delta_{i1}}(\partial
_{\theta})_{i}\tilde{\Lambda}.
\end{equation}
The above Hopf structures then induce the operations (for the details see Ref.
\cite{SW04})
\begin{equation}
f(\theta^{1}\oplus_{L}\psi^{1})=f(\theta^{1}\oplus_{\bar{L}}\psi
^{1})=f^{\prime}+f_{1}(\theta^{1}+\psi^{1}),
\end{equation}
and%
\begin{equation}
f(\ominus_{L}\,\theta^{1})=f(\ominus_{\bar{L}}\,\theta^{1})=f^{\prime}%
-f_{1}\theta^{1}.
\end{equation}

For the sake of completeness let us note that from the coproducts in
(\ref{HopAntBra1}) and (\ref{HopAntBra2}) we can read off the explicit form of
the L-matrices for the Grassmann variables $\theta^{i},$ $i=0,1$. As soon as
we know the action of the scaling operator $\tilde{\Lambda}$ on a given
element $w$ the L-matrices provide a simple method to calculate the braiding
of q-deformed Grassmann variables with $w$ (see also Ref. \cite{MSW04}).

In complete analogy to the symmetrized braided line we can introduce dual
pairings for the antisymmetrized braided line. For normally ordered monomials
we find as non-vanishing pairings
\begin{align}
\big \langle(\partial_{\theta})_{i},\theta^{j}\big \rangle_{L,\bar{R}}  &
=\delta_{i}^{j},\nonumber\\
\big \langle(\hat{\partial}_{\theta})_{i},\theta^{j}\big \rangle_{\bar{L},R}
&  =\delta_{i}^{j},\\[0.1in]
\big \langle\theta^{j},(\partial_{\theta})_{i}\big \rangle_{L,\bar{R}}  &
=-\delta_{i}^{j},\nonumber\\
\big \langle\theta^{j},(\hat{\partial}_{\theta})_{i}\big \rangle_{\bar{L},R}
&  =-\delta_{i}^{j},
\end{align}
and%
\begin{align}
\big \langle(\partial_{\theta})_{0}(\partial_{\theta})_{1},\theta^{1}%
\theta^{0}\big \rangle_{L,\bar{R}}  &  =1,\nonumber\\
\big \langle(\hat{\partial}_{\theta})_{1}(\hat{\partial}_{\theta})_{0}%
,\theta^{0}\theta^{1}\big \rangle_{\bar{L},R}  &  =1,\\[0.1in]
\big \langle\theta^{0}\theta^{1},(\partial_{\theta})_{1}(\partial_{\theta
})_{0}\big \rangle_{L,\bar{R}}  &  =1,\nonumber\\
\big \langle\theta^{1}\theta^{0},(\hat{\partial}_{\theta})_{0}(\hat{\partial
}_{\theta})_{1}\big \rangle_{\bar{L},R}  &  =1.
\end{align}

On the subspace spanned by $\theta^{1}$ these pairings correspond to the
exponentials%
\begin{align}
\exp(\theta^{1}|(\partial_{\theta})_{1})_{\bar{R},L}  &  =1+\theta^{1}%
\otimes(\partial_{\theta})_{1},\nonumber\\
\exp(\theta^{1}|(\hat{\partial}_{\theta})_{1})_{R,\bar{L}}  &  =1+\theta
^{1}\otimes(\hat{\partial}_{\theta})_{1},\\[0.1in]
\exp((\partial_{\theta})_{1}|\theta^{1})_{\bar{R},L}  &  =1-(\partial_{\theta
})_{1}\otimes\theta^{1},\nonumber\\
\exp((\hat{\partial}_{\theta})_{1}|\theta^{1})_{R,\bar{L}}  &  =1-(\hat
{\partial}_{\theta})_{1}\otimes\theta^{1},
\end{align}
which, in turn, give rise to the q-deformed delta functions%
\begin{align}
\delta_{L}^{1}(\eta_{1})  &  =\int d_{L}\theta^{1}\,\exp(\theta^{1}|\eta
_{1})_{\bar{R},L}=\eta_{1},\nonumber\\
\delta_{\bar{L}}^{1}(\eta_{1})  &  =\int d_{\bar{L}}\theta^{1}\,\exp
(\theta^{1}|\eta_{1})_{R,\bar{L}}=\eta_{1},\\
\delta_{R}^{1}(\eta_{1})  &  =\int d_{R}\theta^{1}\,\exp(\eta_{1}|\theta
^{1})_{R,\bar{L}}=\eta_{1},\nonumber\\
\delta_{\bar{R}}^{1}(\eta_{1})  &  =\int d_{\bar{R}}\theta^{1}\,\exp(\eta
_{1}|\theta^{1})_{\bar{R},L}=\eta_{1}.
\end{align}
\vspace{0.16in}

\noindent\textbf{Acknowledgements}

First of all I am very grateful to Eberhard Zeidler for very interesting and
useful discussions, special interest in my work and financial support.
Furthermore, I would like to thank Alexander Schmidt for useful discussions
and his steady support. Finally, I thank Dieter L\"{u}st for kind hospitality.

\end{document}